\documentclass[12pt]{iopart}
\usepackage{array, booktabs, color, eurosym, graphicx, longtable, pbox, pdfpages, url, verbatim}
\usepackage[colorlinks=true,citecolor=blue,filecolor=blue,linkcolor=blue,urlcolor=blue]{hyperref}
\usepackage[alsoload=hep, alsoload=astro, range-phrase=--, range-units=single, separate-uncertainty=true]{siunitx}
\usepackage[utf8]{inputenc}
\usepackage[round,sort]{natbib}
\renewcommand{\cite}[1]{\citep{#1}}
\graphicspath{{figures/}}

\DeclareSIUnit\atm{atm}
\DeclareSIUnit\erg{erg}
\DeclareSIUnit\solarmass{\ensuremath{M_\odot}}  
\DeclareSIUnit\mwe{mwe}
\DeclareSIUnit\year{yr}

\newcommand{\1}[1]{\, \mathrm{#1}} 

\newcommand{\degree}{{}^{\circ}}

\newcommand{\cevns}{CE$\nu$Ns}

\newcommand{\doi}[1]{\href{http://dx.doi.org/#1}{\texttt{doi:#1}}}

\begin{document}

\title{SNEWS~2.0: A Next-Generation SuperNova Early Warning System for Multi-messenger Astronomy}
  
\author{%
S.~Al~Kharusi$^{1}$,
S.~Y.~BenZvi$^{2}$,
J.~S.~Bobowski$^{3}$,
W.~Bonivento$^{4}$,
V.~Brdar$^{5,6,7}$,
T.~Brunner$^{1,8}$,
E.~Caden$^{9}$,
M.~Clark$^{10}$,
A.~Coleiro$^{11}$,
M.~Colomer-Molla$^{11,12}$,
J.~I.~Crespo-Anad\'on$^{13}$,
A.~Depoian$^{10}$,
D.~Dornic$^{14}$,
V.~Fischer$^{15}$,
D.~Franco$^{11}$,
W.~Fulgione$^{16}$,
A.~Gallo~Rosso$^{17}$,
M.~Geske$^{18}$,
S.~Griswold$^{2}$,
M.~Gromov$^{19,20}$,
D.~Haggard$^{1}$,
A.~Habig$^{21}$,
O.~Halim$^{22}$,
A.~Higuera$^{23}$,
R.~Hill$^{17}$,
S.~Horiuchi$^{24}$,
K.~Ishidoshiro$^{25}$,
C.~Kato$^{26}$,
E.~Katsavounidis$^{27}$,
D.~Khaitan$^{2}$,
J.~P.~Kneller$^{28}$,
A.~Kopec$^{10}$,
V.~Kulikovskiy$^{29}$,
M.~Lai$^{30,31}$,
M.~Lamoureux$^{32}$,
R.~F.~Lang$^{10}$,
H.~L.~Li$^{33}$,
M.~Lincetto$^{14}$,
C.~Lunardini$^{34}$,
J.~Migenda$^{35}$,
D.~Milisavljevic$^{10}$,
M.~E.~McCarthy$^{2}$,
E.~O'Connor$^{36}$,
E.~O'Sullivan$^{37}$,
G.~Pagliaroli$^{38}$,
D.~Patel$^{39}$,
R.~Peres$^{40}$,
B.~W.~Pointon$^{41,8}$,
J.~Qin$^{10}$,
N.~Raj$^{8}$,
A.~Renshaw$^{42}$,
A.~Roeth$^{43}$,
J.~Rumleskie$^{17}$,
K.~Scholberg$^{43}$,
A.~Sheshukov$^{20}$,
T.~Sonley$^{9}$,
M.~Strait$^{44}$,
V.~Takhistov$^{45}$,
I.~Tamborra$^{46}$,
J.~Tseng$^{47}$,
C.D.~Tunnell$^{23}$,
J.~Vasel$^{48}$,
C.~F.~Vigorito$^{49}$,
B.~Viren$^{50}$,
C.~J.~Virtue$^{17}$,
J.~S.~Wang$^{47}$,
L.~J.~Wen$^{33}$,
L.~Winslow$^{27}$,
F.~L.~H.~Wolfs$^{2}$,
X.~J.~Xu$^{7}$
and
Y.~Xu$^{23}$
}

\address{$^{1}$~Department of Physics, McGill University, Montr\'eal, QC H3A 2T8, Canada}
\address{$^{2}$~Department of Physics and Astronomy, University of Rochester, Rochester, NY 14627, USA}
\address{$^{3}$~Physics, University of British Columbia, Kelowna, BC V1V 1V7, Canada}
\address{$^{4}$~INFN sezione di Cagliari Istituto Nazionale, Complesso Universitario di Monserrato - S.P. per Sestu Km 0.700, I-09042 Monserrato (Cagliari), Italy}
\address{$^{5}$~Fermi National Accelerator Laboratory, Batavia, IL, 60510, USA}
\address{$^{6}$~Northwestern University, Department of Physics \& Astronomy, 2145 Sheridan Road, Evanston, IL 60208, USA}
\address{$^{7}$~Max-Planck-Institut f\"ur Kernphysik, Postfach 103980, D-69029 Heidelberg, Germany}
\address{$^{8}$~TRIUMF, Vancouver, BC V6T 2A3, Canada}
\address{$^{9}$~SNOLAB, Creighton Mine \#9, 1039 Regional Road 24, Sudbury ON P3Y 1N2, Canada}
\address{$^{10}$~Department of Physics and Astronomy, Purdue University, West Lafayette, IN 47907, USA}
\address{$^{11}$~Université de Paris, CNRS, AstroParticule et Cosmologie, F-75013, Paris, France}
\address{$^{12}$~Instituto de F\'isica Corpuscular (CSIC - Universitat de Val\`encia) c/ Catedr\'atico Jos\'e Beltr\'an, 2 E-46980 Paterna, Valencia, Spain}
\address{$^{13}$~Department of Physics, Columbia University, New York, NY 10027, USA}
\address{$^{14}$~Aix Marseille Univ, CNRS/IN2P3, CPPM, Marseille, France}
\address{$^{15}$~Department of Physics, University of California at Davis, Davis, CA 95616, U.S.A.}
\address{$^{16}$~OATo  Torino \& INFN sezione dei Laboratori Nazionali del Gran Sasso (LNGS), Assergi, Italy}
\address{$^{17}$~Department of Physics, Laurentian University, Sudbury ON P3E 2C6, Canada}
\address{$^{18}$~Department of Physics, Gonzaga University, Spokane, WA 99258, USA}
\address{$^{19}$~Lomonosov Moscow State University Skobeltsyn Institute of Nuclear Physics, 119234 Moscow, Russia}
\address{$^{20}$~Joint Institute for Nuclear Research, 141980 Dubna, Russia}
\address{$^{21}$~Department of Physics and Astronomy, University of Minnesota Duluth, Duluth, MN, 55812, USA}
\address{$^{22}$~Dipartimento di Fisica, Universit`a di Trieste, \& INFN sezione di Trieste, I-34127 Trieste, Italy}
\address{$^{23}$~Departments of Physics, Astronomy, and Computer Science, Rice University, 6100 Main St, Houston, TX, 77005, USA}
\address{$^{24}$~Center for Neutrino Physics, Department of Physics, Virginia Tech, Blacksburg, VA 24061, USA}
\address{$^{25}$~Research Centre for Neutrino Science, Tohoku University, Sendai 980-8578, Japan}
\address{$^{26}$~Department of Aerospace Engineering, Tohoku University, Sendai 980-8579, Japan}
\address{$^{27}$~Massachusetts Institute of Technology, Cambridge, MA 02139, USA}
\address{$^{28}$~Department of Physics, NC State University, Raleigh, NC 27695, USA}
\address{$^{29}$~INFN Sezione di Genova, Via Dodecaneso 33, Genova, 16146 Italy}
\address{$^{30}$~Department of Physics, Cagliari University, Cagliari, CA 09127, Italy}
\address{$^{31}$~Istituto Nazionale di Fisica Nucleare INFN, Italy}
\address{$^{32}$~INFN Sezione di Padova \& Universit\`a di Padova, Dipartimento di Fisica, Padova, Italy}
\address{$^{33}$~Institute of High Energy Physics, Chinese Academy of Sciences, Beijing 100049, China}
\address{$^{34}$~Department of Physics, Arizona State University, Tempe, AZ 85287-1504, USA}
\address{$^{35}$~Department of Physics, King’s College London, London WC2R 2LS, United Kingdom}
\address{$^{36}$~Department of Astronomy and The Oskar Klein Centre, Stockholm University, AlbaNova, 109 61, Stockholm, Sweden}
\address{$^{37}$~Deptartment of Physics and Astronomy, Uppsala University, Box 516, S-75120 Uppsala, Sweden}
\address{$^{38}$~Gran Sasso Science Institute (GSSI) \& INFN sezione di Laboratori Nazionali del Gran Sasso (LNGS), Assergi, Italy}
\address{$^{39}$~Department of Physics, University of Regina, Regina, SK S4S 0A2, Canada}
\address{$^{40}$~Physik-Institut, Universit{\"a}t Z{\"u}rich, Z{\"u}rich, Switzerland}
\address{$^{41}$~Department of Physics, British Columbia Institute of Technology, Burnaby, BC, V5G 3H2, Canada}
\address{$^{42}$~Department of Physics, University of Houston, Houston, TX 77204, USA}
\address{$^{43}$~Department of Physics, Duke University, Durham, NC 27708, USA}
\address{$^{44}$~School of Physics and Astronomy, University of Minnesota Twin Cities, Minneapolis, Minnesota 55455, USA}
\address{$^{45}$~Department of Physics, University of California Los Angeles, Los Angeles, U.S.A.}
\address{$^{46}$~Niels Bohr International Academy and DARK, Niels Bohr Institute, Blegdamsvej 17, 2100 Copenhagen, Denmark}
\address{$^{47}$~Oxford University, Denys Wilkinson Building, Keble Road, Oxford OX1 3RH, UK}
\address{$^{48}$~Department of Physics, Indiana University, Bloomington, IN 47405, USA}
\address{$^{49}$~Department of Physics, University of Torino \& INFN, via Pietro Giuria 1, 10125 Torino, Italy}
\address{$^{50}$~Physics Department, Brookhaven National Laboratory, Upton, NY 11973, USA}

\newpage

\begin{abstract}
The next core-collapse supernova in the Milky Way or its satellites will represent a once-in-a-generation opportunity to obtain detailed information about the explosion of a star and provide significant scientific insight for a variety of fields because of the extreme conditions found within. Supernovae in our galaxy are not only rare on a human timescale but also happen at unscheduled times, so it is crucial to be ready and use all available instruments to capture all possible information from the event. The first indication of a potential stellar explosion will be the arrival of a bright burst of neutrinos. Its observation by multiple detectors worldwide can provide an early warning for the subsequent electromagnetic fireworks, as well as signal to other detectors with significant backgrounds so they can store their recent data. The Supernova Early Warning System (SNEWS) has been operating as a simple coincidence between neutrino experiments in automated mode since 2005. In the current era of multi-messenger astronomy there are new opportunities for SNEWS to optimize sensitivity to science from the next Galactic supernova beyond the simple early alert. This document is the product of a workshop in June 2019 towards design of SNEWS~2.0, an upgraded SNEWS with enhanced capabilities exploiting the unique advantages of prompt neutrino detection to maximize the science gained from such a valuable event.
\end{abstract}

\submitto{\NJP}
\maketitle

\tableofcontents

\newpage

\section{Introduction}
\label{sec:Introduction}

The explosion of a star within the Milky Way Galaxy will provide us with a front row seat of physics under conditions that could never be produced in a terrestrial experiment. While the remnants of the explosion will be observable for many thousands of years, the information about what occurred in the core of the star to cause the explosion will be most easily found in the first tens of seconds. It is therefore imperative that we be able to detect the supernova as soon as it begins --- if not sooner. 

The familiar blast of light visible across the universe is not really when a supernova begins: that electromagnetic radiation starts when the shockwave from the stellar core's collapse reaches the surface and breaks out. Neutrinos are produced at the start of the core collapse process, and escape a supernova explosion well before the photon emission is visible and thus provide the earliest  opportunity to anticipate the imminent appearance of a galactic supernova in time to alert observatories. The Supernova Early Warning System (SNEWS) is an open, public alert system that has provided the capability for such an early warning since 2005 by combining the detection capabilities of a variety of neutrino detectors worldwide~\cite{Antonioli:2004zb}. If several detectors report a potential supernova within a small time window, SNEWS will issue an alert to its subscribers which include astronomical observatories, neutrino detectors, and amateur astronomers and citizen scientists. SNEWS is one of the few successful examples of cyberinfrastructure spanning major neutrino experiments. 

Since the SNEWS network was first established over a decade ago, the particle astrophysics landscape has evolved considerably. The detection of gravitational waves by LIGO/Virgo along with electromagnetic observations of a neutron star merger~\cite{Abbott17}, and the subsequent possible observation of neutrinos from an active blazar by IceCube~\cite{IceCube2018} have ushered in a new era of multi-messenger astrophysics. At the same time, neutrino detector technologies and data analysis techniques have progressed in recent years, and the ability of detectors to detect and analyze neutrinos from galactic supernovae in real time has improved substantially.

In its current form, SNEWS is designed to send a prompt alert based on a simple coincidence, but its functionality can be extended to take advantage of these recent advances. The overarching aim is to enhance the overall science obtained from the next galactic core-collapse supernova (CCSN). These are expected to occur rarely enough ($1.63 \pm 0.46$/century in the Milky Way~\cite{Rozwadowska:2021lll}) that we need to extract all the information possible from the next such CCSN to happen.   Specifically, the goals of SNEWS~2.0 are to:
\begin{itemize}
    \item reduce the threshold for generating alerts in order to gain sensitivity;
    \item reduce alert latency;
    \item combine pointing information from individual experiments and enhance it via timing triangulation;
    \item implement a pre-supernova alert based on the rising neutrino flux which precedes core-collapse;
    \item develop a follow-up observing strategy to prepare the astronomical community for the next galactic supernova; and
    \item engage amateur astronomer and citizen science communities through alert dissemination and outreach.
\end{itemize}

In this paper, we describe enhancements to SNEWS to exploit new opportunities in the era of multi-messenger astrophysics as a means of realizing these goals.  Section~\ref{sec:Introduction} introduces the existing network and the overall plan. Section~\ref{sec:CCSNe} provides background on the types of transient events that are of interest to SNEWS and the characteristics of their signals at Earth. Section~\ref{sec:pointing} describes how pointing information can be extracted from a neutrino signal via anisotropic interactions and signal triangulation between multiple detectors. Section~\ref{sec:preSN} explores the possibility of producing an earlier warning by measuring the pre-supernova neutrino flux from stars during silicon burning, which directly precedes core-collapse. Section~\ref{sec:snews} is a review of the SNEWS design, how SNEWS~2.0 alerts will be disseminated, and how follow-up observations can be incorporated, while Section~\ref{sec:detectors} describes the experiments involved. Finally, Section~\ref{sec:outreach} discusses public outreach, including how SNEWS~2.0 will interface with amateur astronomers and citizen scientists.

\subsection{Current Configuration (SNEWS~1.0)}
\label{sec:current_config}

The SNEWS~1.0 system was designed to give a supernova neutrino alert that was
\begin{itemize}
    \item \textbf{Prompt}, providing an alert within minutes and followup within hours; 
    \item \textbf{Positive}, with less than one false alarm per century; and
    \item \textbf{Pointing}, providing a sky location if and when possible by passing along experiments' estimates.
\end{itemize}
The design was primarily driven by the Positive requirement. Only extremely high quality coincidences could automatically trigger an alert. All other coincidences would require human intervention before an alert could be triggered. 

SNEWS (\url{https://snews.bnl.gov}) currently involves an international collaboration of supernova neutrino detectors: Super-Kamiokande, LVD, IceCube, Borexino, KamLAND, HALO, and Daya Bay (with NOvA, KM3NET, and Baksan testing their connections to join soon). SNEWS has been operational since 1998 and has been running in a fully-automated mode since 2005 with near-100\% up-time. The main idea of SNEWS is to provide \textit{prompt, high-confidence} alerts of nearby CCSN by requiring a burst coincidence between detectors; this allows alarms from individual detectors to go out promptly without needing a human check.

SNEWS operates two ``coincidence servers'': a primary server at Brookhaven National Laboratory and a backup at the University of Bologna.  The participating experiments each run their own online supernova monitors, and run client code provided by SNEWS to send datagrams to the servers if supernova-like bursts are observed.  The minimal information provided in the client datagram is the experiment, time of the first event of the burst, and a burst-quality parameter. Experiments may choose to provide directional and burst size information if it is available promptly. ``Gold"  alerts, for which input datagrams must satisfy several quality criteria, are sent out automatically by the server to a mailing list if a coincidence within \SI{10}{\second} is found. Email alerts are provided first to the other experiments and to  ``express-line" subscribers (LIGO, ANTARES, and the Gamma-Ray Coordination Network (GCN)\cite{bacodine}), and they are also available by direct socket connection (NOvA and XENON1T). ``Silver" alerts are sent to the experiments only.

The current SNEWS requirement for accidental-coincidence alerts is that they must occur less than once per century.  If individual input alarm rates become too high, or there are other low-quality indicators in the input, coincidence output is demoted to ``silver''.  Single experiments may also send datagrams to SNEWS with sufficiently well-vetted alarms to be propagated automatically as ``individual'' alerts. Detailed information on coincidence criteria for the configuration of SNEWS can be found in~\cite{Antonioli:2004zb, Scholberg:2008fa}. A weekly test alert is sent via GCN every Tuesday at noon Eastern.

\subsection{SNEWS~2.0}

SNEWS~2.0 will be an upgrade of the SNEWS system for the age of multi-messenger astronomy. In this environment, false alarms are acceptable, low probability events should be reported, and SNEWS will be one of many multi-messenger alert systems. Nevertheless, SNEWS remains unique by combining data from different neutrino observatories, and providing clear summaries of neutrino data for astronomers. This section outlines a number of improvements which will be made using more detectors and different techniques than the original SNEWS.

Recent additions to the suite of potential detectors for the next Galactic supernova are large dark matter detectors. Capitalizing on the recently-discovered coherent elastic neutrino-nucleus scattering (CE$\nu$NS)~\cite{Akimov:2017ade}, those detectors rely on the coherent enhancement of the neutrino cross-section for supernova burst detection that can be probed thanks to low ($<\mathrm{keV}$) energy thresholds. Those detectors provide a flavor-insensitive detection channel and thus a total-flux measurement of the total energy going into neutrinos, independent of, for example, uncertainties from neutrino oscillations~\cite{Lang:2016zhv,Chakraborty:2013zua}. Furthermore, the combination of CE$\nu$NS and inelastic interaction channels will help to disentangle oscillation effects.  Currently, these detectors need improved understanding of backgrounds at the lowest energies that are relevant here~\cite{Aprile:2013blg,Sorensen2017ElectronTB,Sorensen:2017kpl}.  So far, XENON1T has a dedicated trigger following a SNEWS alert.  SNEWS~2.0 aims to go a step further to enable the integration of dark-matter detectors as inputs to SNEWS.

SNEWS~2.0 also intends to develop and provide a \textit{true} pre-supernova alert---a pre-core-collapse alert. This is based on the predicted uptick in neutrino production that accompanies the final burning stages of a doomed star~\cite{Odrzywolek:2003vn, Kato:2017ehj, Patton:2015sqt}. This alert has been implemented in KamLAND~\cite{Asakura:2015bga}, and provides a $\SI{3}{\sigma}$ detection 48 hours prior to the explosion of a \SI{25}{\solarmass} star at \SI{690}{\parsec}. Extending this alert to the network should expand the sensitivity to a larger fraction of the galaxy.

SNEWS~2.0 could also be used to communicate and organize planned shutdowns or downtime in each detector to ensure that the overall supernova detection livetime is not affected, since other participating experiments can cover.

\subsection{Lowering the Threshold}
\label{sec:lower_threshold}

One general benefit of combining detectors' data real-time would be the lowering of the effective threshold for observing a signal. An astrophysical neutrino signal would be observable in many detectors at once, but might be not strong enough to be significant in any one detector.  Since most detectors are large enough to be sensitive to a CCSN somewhere in our galaxy, the most obvious benefit of being able to see farther against the inverse-square law flux suppression with distance is not as useful as it might seem at first, with the exception of a supernova in the Magellanic clouds, where the distance is substantial and the flux is borderline for most detectors. However, a combination of detector signals allowing for more sensitivity to low flux would enhance the world's ability to notice any unusually low flux events.  Four important examples of this are the pre-supernova neutrinos introduced in the previous section (elaborated on in Sec.~\ref{sec:preSN}), the neutrinos emitted at the latest times in the burst~\cite{2020arXiv200804340W}, and neutrinos from three other types of potential transient bursts described in Sec.~\ref{sec:transients}.  As the current range of such detection is only hundreds of parsecs, increasing sensitivity via comparing sub-threshold signals in different experiments will increase the number of progenitors under observation by a factor of distance cubed.

These pre-supernova neutrinos have a lower energy ($\sim$ a few MeV) and a much lower flux than core-collapse supernova neutrinos. Being low energy, there are also more potential background events to confuse with a potential signal. In the near future, lowering the energy thresholds and better background rejection are in the plans for both running and planned detectors, expanding the experiments able to do so (currently, only KAMLAND has this capability). For example, delayed coincidence with gadolinium is being implemented Super-Kamiokande, which effectively lowers the energy threshold by the confirmation of inverse-$\beta$ decay events~\cite{Beacom:2004}. At liquid-scintillator detectors, the low energy threshold (\SI{1.8}{\mega\electronvolt}) is achieved via good scintillation light production. The detection of keV-neutrinos will be practical via CE$\nu$NS at large dark-matter detectors. A supernova alert with pre-supernova neutrinos has been investigated in several recent works \citep{Asakura:2015bga,Kato:2017ehj,Raj:2019wpy,Simpson:2019xwo,Li:2020gaz}.

In addition to the developments in individual neutrino detectors, their combination via SNEWS 2.0 reduces the uncertainty of the supernova alert and effectively lowers the threshold for the alert issue. Because the energy of pre-supernova neutrinos increases as the pre-supernova stellar core evolves, an earlier alert is possible by combining different experiments' sub-threshold data to provide plenty of preparation time for the detection of other observables. This earlier alert will maximize the information to be gained from multi-messenger astronomy, yielding information of supernovae from a different perspective. Moreover, pre-supernova neutrinos will be one of the useful tools to prove the theory of stellar evolution \citep{Kato:2015,Yoshida:2016imf} and long term detection over several stellar evolutionary phases and experiments will improve the results.

\section{Stellar Core Collapse Signals}
\label{sec:CCSNe}

\begin{figure}[htb]
\centering
\includegraphics[scale=0.4]{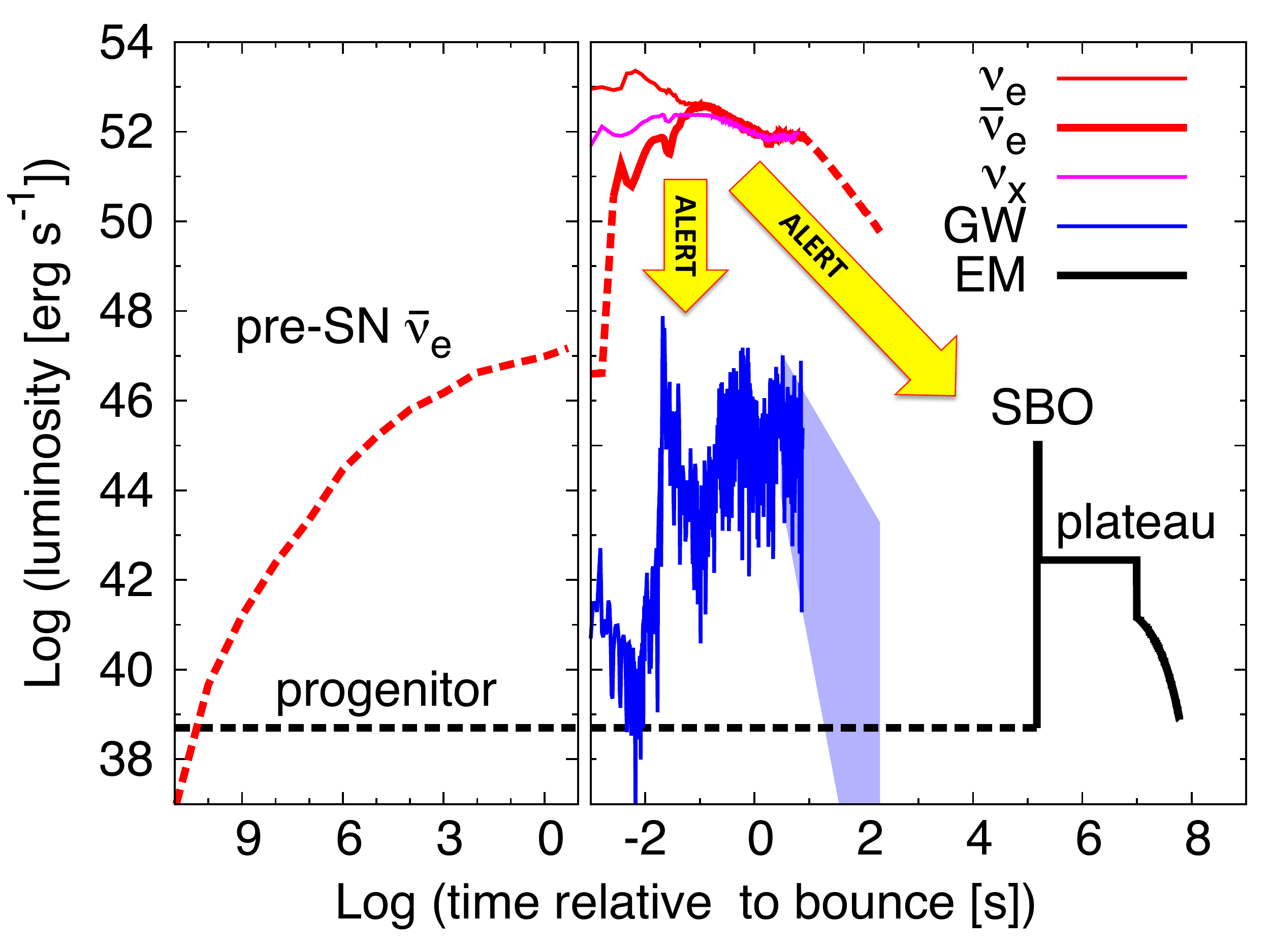}
\caption{Time sequence for multi-messenger signals pre- (left panel) and post- (right panel) core collapse of a non-rotating \SI{17}{\solarmass} progenitor star. Neutrinos ($\nu_e$, $\bar{\nu}_e$, and $\nu_x$ are shown by red, thick red, and magenta lines, respectively, where $\nu_x$ represents heavy-lepton neutrinos: $\nu_\mu$, $\nu_\tau$, $\bar{\nu}_\mu$, and $\bar{\nu}_\tau$), gravitational waves (blue line), and electromagnetic signals (black line) are shown. Solid lines are predictions from a hydro-dynamical simulation with axis-symmetric radiation, while dashed lines are approximate predictions. Neutrino emission prior to collapse arises from the last moments of stellar evolution, but is quickly overtaken during collapse by the neutrino burst. The electromagnetic signal exhibits the shock breakout (SBO), plateau, and decay components. Note that the height of the curves does not reflect the energy output in each messenger; the total energy emitted after the bounce in the form of $\bar{\nu}_e$, photons, and gravitational waves are $\sim$\SI{6e52}{\erg}, $\sim$\SI{4e49}{\erg}, and $\sim$\SI{7e46}{\erg}, respectively. The focus of SNEWS~2.0 is to establish the neutrino burst as an alert for gravitational waves and electromagnetic followup, as shown by arrows.
Adapted from \cite{Nakamura:2016kkl}.}
\label{fig:multimessengersignal}
\end{figure}

The dominant source of supernova neutrino bursts are Core Collapse SuperNovae (CCSN). These type of supernova occur when a massive (more than $\sim$\SIrange{8}{10}{\solarmass}) star, after successively burning elements from hydrogen to silicon, forms an inert, but growing, iron core. This core soon reaches the effective Chandrasekhar mass and collapses due to the unmatchable strength of gravity.  The collapse continues until the density reaches nuclear densities where the equation of state stiffens and the nuclear force is able to stabilize the core against gravity. The formation of the protoneutron star also leads to the formation of a shock wave.  It is this shock wave that for successful supernovae will traverse the star over the course of minutes to hours and unbind all but the innermost material. Core-collapse supernovae emit signals in three cosmic messengers---electromagnetic emission, neutrinos, and gravitational waves---and also cosmic rays at later times. Figure \ref{fig:multimessengersignal} shows an example of the expected time sequence for these signals, starting from before (left panel) and after (right panel) core bounce. 

While each cosmic messenger is valuable by itself, when analyzed together, they provide a comprehensive understanding that is impossible to achieve from any single one of them alone. Multi-messenger astrophysics had two foundational discoveries in 2017: a binary neutron star merger that produced gravitational waves and electromagnetic radiation \cite{Abbott17}, and the coincident detection of high-energy neutrinos and electromagnetic emission from a blazar \cite{IceCube2018}. We expect multi-messenger observations of the next Galactic core-collapse supernova offer similar synthesis opportunities. Coordinating timely follow-up observations that enable a true multi-messenger analysis demands rapid identification and characterization of the neutrino signal, along with prompt broadcasting to ensure that transient emission only detectable on short time scales is recorded. Enabling this coordination is the SNEWS2.0 raison-d'{\^e}tre.

\subsection{Neutrinos from Core Collapse Supernovae}

The neutrino emission from a core collapse supernova in our Galaxy cannot be hidden in any way. The neutrinos are not obscured by dust as electromagnetic signals may be, nor would failure of the explosion mean the supernova would evade our detection: a large burst of neutrinos would still be emitted prior the formation of a black hole. Finally, the present detection horizon for neutrinos reaches out beyond the edge of the Milky Way. For all these reasons, neutrinos are a unique messenger to provide a compelling trigger for an alert. Coupled with gravitational waves (whose detection will also be enhanced by the precise timing information provided by neutrinos) and electromagnetic observations, the neutrinos will allow us to piece together a comprehensive picture of the supernova from the moment of core collapse to supernova shock breakout and beyond. Expected features in the neutrino signal will permit us to probe a long list of topics, including: key aspects of the supernova explosion mechanism ({\it e.g.}, fluid instabilities vs.\ core rotation), the nuclear equation of state, the stellar radius and interior structure, explosive nucleosynthesis, the nature of the remnant core (neutron star vs.\ black hole), as well as answer questions about the fundamental properties of neutrinos, and even test Beyond-Standard-Model physics \cite{Horiuchi:2017sku}. To fully develop these prospects, it is essential the supernova be detected to the latest times possible with good flavor and spectrum information~\cite{Nakazato:2020ogl,2020arXiv200804340W}.  The multi-messenger nature of the supernova signal greatly helps in extracting this information from the neutrinos. For example, it has been shown by \cite{Warren:2019lgb} that the neutrino emission is correlated with the gravitational wave signal which would aid in disentangling the neutrino oscillation effects.

\subsection{Gravitational Wave Signals from Core Collapse}

Together with neutrinos, gravitational waves provide a unique probe of the core collapse in realtime. The emission of gravitational waves is strongly dependent on the asymmetry of the collapsing core and the nuclear equation of state, opening a view of the collapsing core complementary to neutrinos \cite{Janka:2017vcp,Kotake:2011yv,Morozova:2018glm}. By combining gravitational waves with neutrinos and electromagnetic waves, key aspects of the collapse, from the spin of the collapsed core to the supernova explosion mechanism and black hole formation, become be more robustly probed. At present, even for a conservative prediction of the emitted gravitational wave signal, detectors such as Advanced LIGO, Advanced Virgo, and KAGRA are able to detect CCSN gravitational waves out to a few kpc from the Earth, while future detectors such as the Einstein Telescope can reach the entire Milky Way. The detection horizon of the circular polarization can be significantly larger than the gravitational wave amplitude, and can also help reveal inner dynamics \cite{Hayama:2018zgb}. 

Several mechanisms can generate gravitational waves during a CCSN, 
for a
recent review see~\cite{Abdikamalov:2020jzn} and references therein.
The majority of these signals have the common feature of being short and ``burst like'', {\it i.e.} impulsive signals lasting less than a second and very difficult to model. These characteristics make detection more challenging. The identification of a temporal window in which to look for the signal significantly increases the detection efficiency. Neutrinos can provide the best temporal trigger for this gravitational wave search; indeed the neutrino signal for a Galactic CCSN allows the time of the core ``bounce’’ to be identified within a window of $\sim 10$ ms or less \cite{Pagliaroli:2009qy,Halzen:2009sm}. The use of this information improves the background reduction of gravitational wave detectors, with consequent increases of the detection capability \cite{Nakamura:2016kkl}.  
In the case of long-lasting GW emission due to neutron star
oscillations~\cite{Radice:2018usf} the identification of the time of the bounce through
neutrinos could provide a reference point to start the search.

\subsection{Electromagnetic Signals}

EM radiation in the first hours to days after core collapse explosion provides critical information about the progenitor star and the overall energy budget and dynamics of the core collapse explosion. A few hours to days after the core collapse, the supernova shock breaks out of the progenitor surface, suddenly releasing the photons behind the shock in a flash bright in UV and X-rays, known as shock breakout (SBO) emission. SBO has been observed on rare occasions in extragalactic systems \cite{Soderberg:2008uh,Gezari2010,Bersten2018}. The SBO signal provides important information about the supernova, such as the radius, mass, and structure of the progenitor star, and the kinematic energy associated with the rapidly expanding ejecta. Initial observations of the gamma flux from the first moments of a SN will be also be important to help constrain the terrestrial effects of gamma rays from historical SNe on atmospheric chemistry and climate science~\cite{Brakenridge:2020,Jull:2018}.  Knowledge of where and when to anticipate the signal will ensure that the peak luminosity and duration of the SBO (strongest at UV and soft X-ray wavelengths) is not lost. Even including SN1987A, the precise time between onset of core collapse and shock break out has never been measured~\cite{SN1987Areview1989,EB1992}.  Prompt alert and coordinated follow up with SNEWS~2.0 will make this possible. 

\subsection{Other Transients}
\label{sec:transients}
While core collapse supernovae are expected to be the dominant type of supernovae in the Milky Way, they are not the only astrophysical sources of neutrino bursts. Bursts are also expected from Type Ia supernovae (SNIae), pair-instability supernovae (PISNe), compact object mergers, and possibly others yet unknown. There are a number of questions, many fundamental, about these other neutrino transients so that a neutrino signal from any one of them would represent as rich an opportunity to advance our knowledge as the signal from a core collapse. 

\subsubsection{Type Ia Supernovae}
\label{sec:typeIa}
The progenitor systems of Type Ia supernovae and their associated explosion mechanisms remain debated. The possible progenitors of SNIae --- and the observational constraints upon the various scenarios --- are discussed extensively in Maoz, Mannucci and Nelemans \cite{Maoz2014} and Ruiz-Lapuente \cite{Ruiz-Lapuente2014}. Even if we accept the canonical model of a SNIa as the disruption of a Chandrasekhar mass (\SI{1.4}{\solarmass}) carbon-oxygen white dwarf, many different scenarios for how the explosion proceeds can be found in the literature \cite{DDTOriginal,Plewa2004,PRDoriginal}. We refer the reader to Hillebrandt \emph{et al.} \cite{Hillebrandt2013} for a review.

The neutrino emission from a limited number of SNIa simulations has been computed \cite{Odrzywolek2011a,2016PhRvD..94b5026W,2017PhRvD..95d3006W}. Wright \emph{et al.} considered the most optimistic case (known as the DDT) and a more general case (their GCD case). The number of events they expect in a \SI{374}{\kilo\tonne} water-Cherenkov detector from a SNIa at a distance of \SI{10}{\kilo\parsec} is of order 1 for the DDT case and 0.01 for the less optimistic GCD. A SNIa would have to be within a few kpc in order to detect tens of events but the probability the next Galactic supernova is within \SI{5}{\kilo\parsec} is only of order 10\% according to Adams \emph{et al.}~\cite{Adams:2013ana}. 

\subsubsection{Pair Instability Supernovae}
\label{sec:PISN}

Very massive stars can explode as a PISN if they form a carbon-oxygen core in the range of \SI{64}{\solarmass} $< M_{\mathrm{CO}} <$ \SI{133}{\solarmass} ~\cite{2002ApJ...567..532H}. The temperatures in these cores are sufficiently high and the electron degeneracy sufficiently low that electron-positron pairs are created. The formation of the pairs softens the equation of state causing a contraction of the core triggering explosive burning of the oxygen \cite{1967PhRvL..18..379B,Ravaky.Shaviv:1967,Fraley1968}. The energy released is enough to unbind the entire star leaving behind no remnant. Some models of PISNe produce very large amounts of $^{56}$Ni and PISN are candidates for some superluminous supernovae \cite{Smith:2006dk,2009Natur.462..624G,2012Natur.491..228C,Lunnan.ea.PS1-14bj:2016}.

The long-standing expectation of theorists is that only metal-free stars could remain sufficiently massive to explode as PISN \cite{2002ApJ...567..532H}. However this expectation has been challenged in recent years. \cite{2007A&A...475L..19L} found PISN can occur in stars with metallicities as large as $Z_{\odot}/3$ while \cite{Georgy.Meynet.ea:2017} obtained the conditions for a PISN at near solar metallicities if they included surface magnetic fields. Thus, theoretically at least, a PISN in the Milky Way or one of its satellites cannot be ruled out. 

The rate of PISNe is uncertain because a) observationally we lack an unambiguous method for discriminating these kind of supernovae from the others and b) theorists have not reached a consensus on which masses at a given metallicity produce these kinds of events. The estimate by Langer~\emph{et al.} is for a rate of \SI{e-4}{\per\year} but that could be larger by an order of magnitude if the recent revisions to the progenitor are correct. 

The neutrino signals from two PISN simulations have been computed by ~\cite{PhysRevD.96.103008}. The two models they considered were a low-mass and high-mass case so that the computed signals spanned the range of possibilities. The flux at Earth from a `small' PISN at \SI{10}{\kilo\parsec} was similar to the most optimistic SNIa case i.e. around 1--2 events, but for a `large' PISN at the same distance the flux was much larger, between 50--100 events depending upon the equation of state and the neutrino mass ordering.

\subsubsection{Compact Object Mergers}

The neutrino emission from merging neutron stars has been computed by \cite{2003MNRAS.342..673R} and the neutrino emission from a black-hole-neutron star merger simulation was computed by \cite{2009PhRvD..80l3004C}. In both cases the neutrino emission is similar to that from a core-collapse supernova {\it i.e.}, the neutrino luminosities and mean energies are within a factor of a a few of those found in core-collapse simulations), and therefore give similar event rates in detectors. However there are differences: in a core-collapse there are more neutrinos than antineutrinos emitted and the duration of the burst is of order \SI{10}{\s}. In a neutron star merger the opposite matter-anti-matter ratio is expected and the signal lasts for 1 second unless the supermassive neutron star can be prevented from forming a black hole. The rate of black-hole - neutron star mergers in not known precisely but the rate of  neutron star-neutron star mergers can be better estimated because there exist a number of such systems in the Milky Way. \cite{Abadie:2010cf} calculate the likely event rate to be \SI{e-4}{\per\year} while \cite{Kalogera_2004a,Kalogera_2004b} give the plausible range to be from \SI{e-6}{\per\year} to \SI{e-3}{\per\year}.

To detect neutrinos from black-hole-neutron star and neutron star-neutron star mergers is challenging. A promising strategy is to search for neutrinos in time-coincidence with detections of mergers in gravitational waves, using a time window of, e.g., \SI{1}{\s} after each merger \cite{Kyutoku:2017wnb,Lin:2019piz}.  This strategy reduces the backgrounds very effectively so that, in fortunate circumstances, even the detection of a single, time-coincident neutrino can be statistically significant. If alerts from Advanced LIGO \cite{Abadie:2010cf} (which has a distance of sensitivity to mergers of about \SI{200}{\mega\parsec}) were used, a megaton water Cherenkov detector could record about 1 neutrino detection per century \cite{Kyutoku:2017wnb}. When operating in synergy with third-generation gravitational wave observatories, like the proposed Einstein Telescope \cite{Punturo:2010zz}, and Cosmic Explorer \cite{Evans:2016mbw} (sensitivity up to redshift $z\sim 2$), the same detector could identify of up to a few neutrinos from mergers per decade, and start placing constraints on the parameters space already after a decade or so of operation \cite{Lin:2019piz}.

\section{Pointing to the Supernova with Neutrinos}
\label{sec:pointing}

\begin{figure}[htb!]
\centering
\includegraphics[scale=1.0]{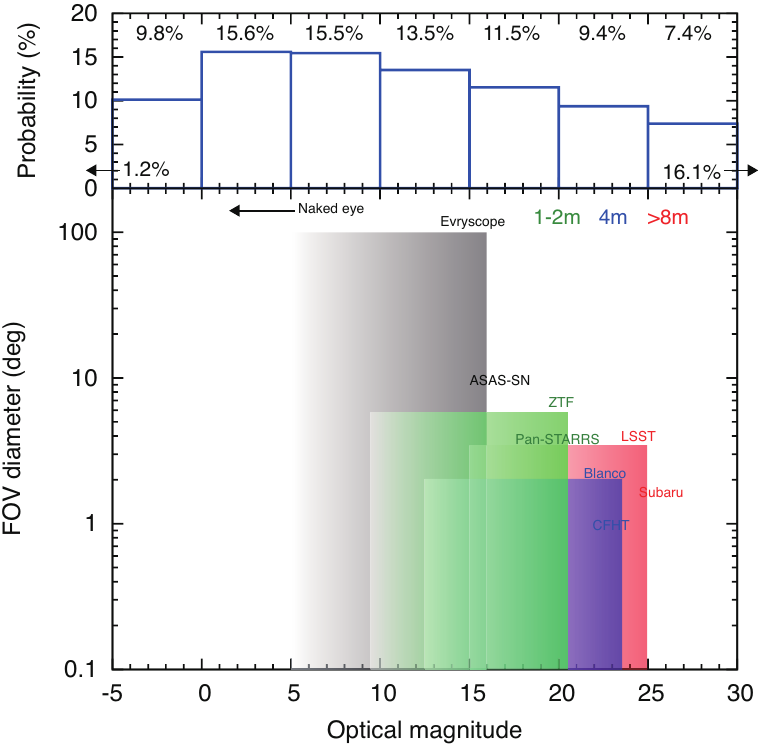}
\caption{Optical follow-up requirements for the next Galactic supernova. The top panel shows a histogram of the apparent magnitude probability distribution for the shock breakout signal of a Galactic supernova (uncertainties not shown; see text). The bottom panel shows the magnitude sensitivity range and fields of view (FOV) of optical telescopes: ASAS-SN, Blanco, CFHT, Evryscope, LSST, Pan-STARRS, Subaru, and ZTF. When the optical magnitude is brighter than $\sim 15$ mag, early detection is feasible thanks to the wide FOV of small-aperture telescopes. However, fainter cases are more challenging since there are no $>1$ m telescope with a FOV larger than $\sim 6^\circ$ diameter. Therefore, this sets the target accuracy for triangulation by SNEWS~2.0. Note that for the brightest supernovae, telescopes will need high-quality filters for accurate photometry, shown by the fading color. Adapted from \cite{Nakamura:2016kkl}.}
\label{fig:followup}
\end{figure}

While supernovae are optically highly luminous, a large fraction are anticipated to be heavily attenuated by dust along the line of sight, typically within the disk of the Milky Way. The supernova optical signal as observed at Earth has been estimated adopting a standard intrinsic supernova luminosity distribution, Galactic distribution for supernova occurrence, and a simple model of Galactic dust extinction. According to~\cite{Adams:2013ana,Nakamura:2016kkl}, the dominant fraction (some 50\%) will be observable with 1--2 meter class telescopes in the optical band. An additional 10\% of supernova can be observed by larger 4--8 meter class telescopes, while the faintest 25\% will be as faint as 25--26 magnitudes and require dedicated observations by the largest available telescopes (however, dust attenuation uncertainties are large in this regime, being at least a few magnitudes). This is illustrated in the top panel of Figure \ref{fig:followup}. The fields of view of the relevant telescopes typically do not cover more than several degrees in a single pointing, as shown by the rectangular boxes on the bottom panel of Figure \ref{fig:followup}. This highlights the quantitative demands for a combined rapid (in time) and accurate (in direction pointing) alert.  

In order to be an effective trigger, the neutrino alert needs to be sent faster than the delay between neutrinos and the first electromagnetic signal, and also alert the pointing in the sky to within a few degrees.  The SNEWS~2.0 alert will be automated and sent electronically to meet the timing demands (see Section~\ref{sec:snews}). Triangulation-based pointing will be implemented in SNEWS~2.0. Such pointing techniques were originally explored in \cite{Beacom:1998fj} and further developed by \cite{Muhlbeier:2013gwa,Brdar:2018zds,Linzer:2019swe}. 

An important consideration is that an alert may be associated with an event that is challenging to observe at EM wavelengths. Possible scenarios include a distant supernova associated with large extinction due to dust and/or formation of a black hole with a weak explosion. In these cases follow up strategies informed by the neutrino alert are needed. For example, in extreme scenarios the Vera Rubin Observatory's field of view and photometric depth are uniquely suited \cite{Walter2019}.

\subsection{Anisotropic Interactions}
\label{sec:anisotropicInteractions}

\subsubsection{Water Cherenkov}

Large water Cherenkov detectors (WCDs), such as Super-K and the future Hyper-K, have potentially good supernova pointing capability through the anisotropic neutrino-electron elastic scattering (ES) interaction \cite{Beacom:1998fj,Tomas:2003xn}. The majority of supernova neutrinos interact in WCDs through inverse beta decay (IBD). In IBD the direction of the outgoing positron is nearly random with respect to the direction of the incoming supernova neutrino. The angular distribution of measured positron directions from a large number of IBD events is necessary to detect (and possibly use) the small anisotropy in the direction of the neutrino flux.

\begin{figure}[ht]
\centering
\includegraphics[width=0.9\textwidth]{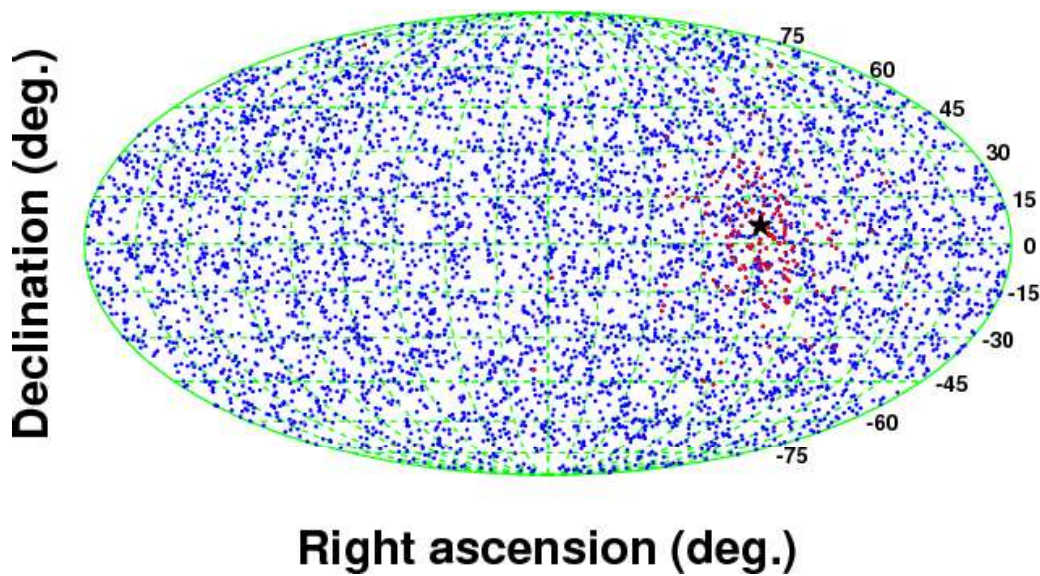}
\caption{Reconstructed skymap of MC simulated Super-K supernova event direction vectors. Red points: anisotropic elastic scatter events, blue points: IBD and other nearly isotropic events, star point: direction vector from supernova~\cite{Abe:2016waf}.}
\label{fig:SK_SN_event_angular_distribution}
\end{figure}

Fortunately, a few percent of the interactions are due to elastic scatter of the supernova neutrinos from electrons. The outgoing electrons are preferentially forward-scattered and the reconstructed direction of the scattered electron is correlated to the direction of the incoming neutrino. The angular distribution of measured electron directions from ES shows a strong correlation with the direction of the neutrino flux. The magnitude of the anisotropy varies with neutrino energy and flavor. However, since the ES cross sections are small, a large detector mass is required to measure enough ES interactions for direction finding. At present WCDs cannot accurately differentiate between IBD and ES interactions, although adding increasing amounts of Gadolinium to Super-K will improve this. Thus, the high ratio of IBD to ES interactions reduces the signal-to-noise ratio of the direction signal, as shown in Figure~\ref{fig:SK_SN_event_angular_distribution}. 

In a large WCD the direction of the neutrino flux (and therefore the supernova direction) may be found by analyzing the  energies and direction vectors of the electrons/positrons, reconstructed from the Cherenkov ring for each event. For example, the current Super-K real-time supernova burst monitor performs SN direction finding using a maximum likelihood method~\cite{Abe:2016waf}. The direction and energy of each event is used to calculate the likelihood of a given supernova direction and event reaction channel based on a probability density function (PDF) determined from supernova neutrino flux models and Super-K MC simulations. The supernova direction angles, and other parameters, are varied until the total likelihood is maximized. The pointing accuracy for a supernova at \SI{10}{\kilo\parsec} is estimated using a modern supernova model to be 4.3--5.9$\degree$ (68.2\% C.L.) covering all combinations of neutrino oscillations and mass orderings.  This will improve to 3.3--4.1$\degree$ for the fully doped SK-Gd.

The accuracy of supernova direction finding based on the anisotropy of ES events depends on the number of events. This varies with detector volume, supernova distance, neutrino oscillations and supernova mass and neutrino emission mechanisms. The coming Mtonne-scale water Cherenkov detectors, such as Hyper-Kamiokande, will also have improved direction finding  due to increased statistics in the ES channel. The pointing accuracy for Hyper-K is expected to be 1--1.3$\degree$ \cite{Abe:2018uyc}.

The introduction of small amounts of gadolinium into the water volume of WCDs will allow accurate tagging of individual IBD events. Thus, direction finding routines could de-weight or exclude the IBD events, potentially increasing the speed and pointing accuracy. Super-K will be implementing such an upgrade in the near future.

\subsubsection{Liquid Argon}
\label{sec:larpoint}
Liquid argon time projection chambers (LArTPC) detectors have the ability to do fine-grained tracking of the final-state particles, and, like water Cherenkov detectors, can exploit the intrinsic directionality of anisotropic interactions in the detector. Ability to tag different interaction channels also helps. Elastic scattering interactions on electrons with well-known energy dependence can be used, as well as the charged-current $\nu_e$ absorption interactions on $^{40}$Ar.  The latter have a relatively weak anisotropy but large statistics.

Unlike water Cherenkov signals, the tracked electrons have a head-tail ambiguity that results in about half of them with a fake reconstructed backwards direction.  This ambiguity can be resolved statistically using sophisticated reconstruction techniques.  Improvement by using the directionality of bremsstrahlung gammas, which are emitted preferentially in the electron travel direction, has been demonstrated in DUNE.  Using a likelihood technique with the ensemble of electron scattering and $\nu_e$CC events, DUNE has demonstrated about 5$^\circ$ pointing for a \SI{10}{\kilo\parsec} supernova signal~\cite{Abi:2020evt}.

\subsubsection{Liquid Scintillator}

Liquid scintillator and water Cherenkov detectors alike are mostly sensitive to IBD interactions -- the major difference between the two being that such interactions are considered a background for supernova pointing in the latter while they are considered a signal in the former. Indeed, while considered isotropic at first order, the positron and neutron emitted after an IBD interaction both possess a slight energy-dependent anisotropy~\cite{Strumia:2003zx}. At the energies of interest for supernova neutrino detection, the positron quickly deposits its energy and therefore most of the anisotropy is carried away by the neutron, always emitted in the forward direction. Although this appears to be similar to the forward emission of an electron in elastic scattering, detecting a neutron direction is arduous in large scintillator detectors. In the vast majority of cases, only the position of the neutron capture vertex after thermalization and diffusion can be determined. For each IBD interaction, a direction vector, starting at the reconstructed positron vertex and ending at the reconstructed neutron capture vertex, can be defined. Due to the smearing caused by the neutron transport after its creation, a single IBD vector is not sufficient to efficiently reconstruct its neutrino incoming direction. However,the analysis of thousands or more of IBD interactions can help reconstruct the statistical direction of an incoming neutrino flux, as demonstrated by the CHOOZ collaboration with about~2,700 events~\cite{Apollonio:1999jg}.

\begin{figure}[ht]
\centering
\includegraphics[width=0.9\textwidth]{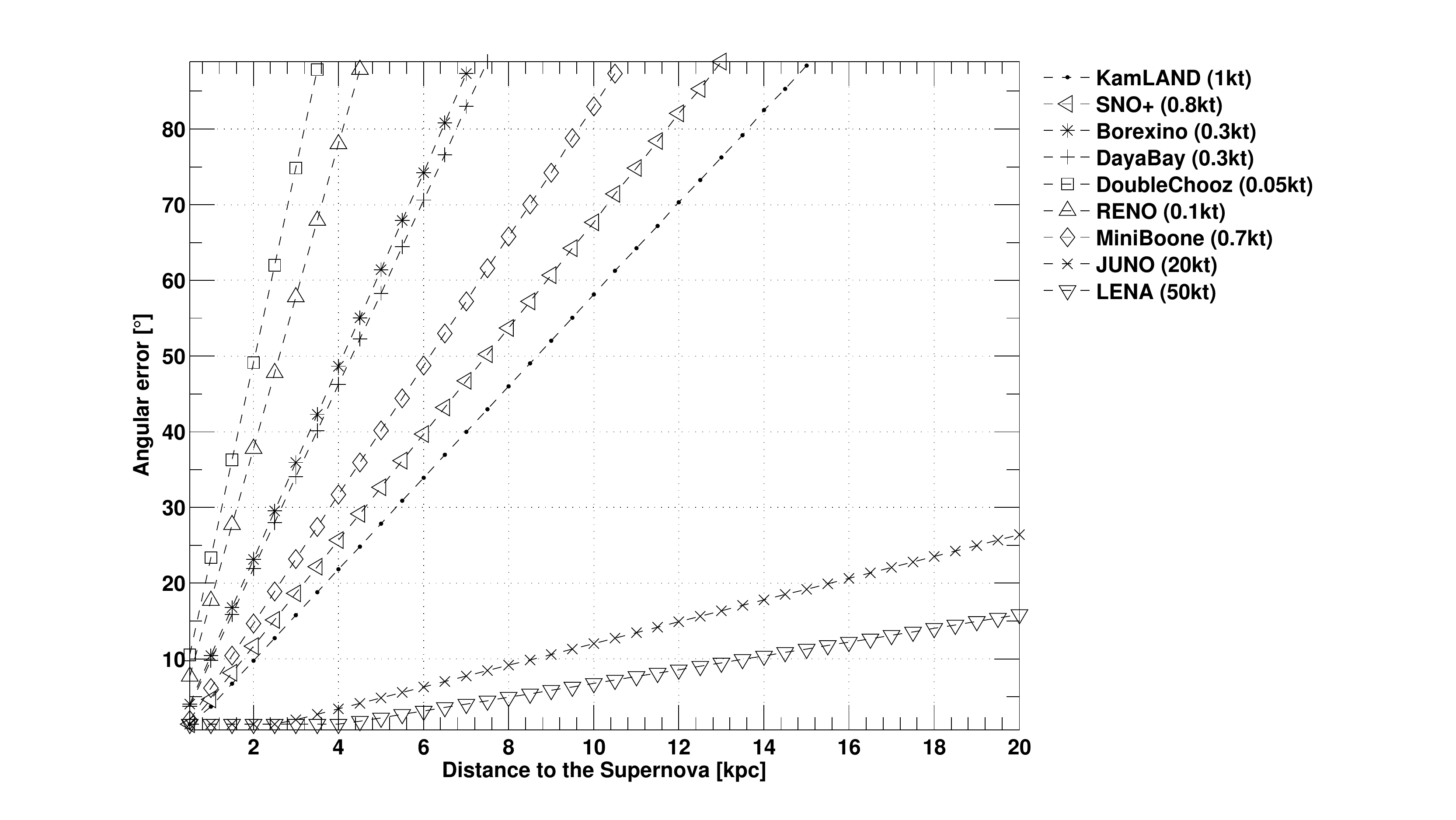}
\caption{Angular uncertainty (68\% C.L.) as a function of a galactic supernova distance for different existing and proposed detectors along, with their associated masses~\cite{Fischer:2015oma}.}
\label{fig:pointing_IBD_all_detectors}
\end{figure}

Such an analysis can be performed to determine the expected direction of a supernova-induced neutrino flux, as shown in~\cite{Fischer:2015oma}. In this study, the statistical nature of the supernova direction reconstruction through IBD anisotropy was exploited by combining the direction vectors of all IBD interactions from several liquid scintillator-based detectors, existing or proposed. While the pointing capabilities of individual existing detectors, shown in Figure~\ref{fig:pointing_IBD_all_detectors}, are no match for the accuracy of Super-K, their combination, as well as the introduction of JUNO in a near future, provides non-negligible pointing information. With the addition of JUNO to the existing large liquid scintillator detectors, supernova pointing accuracy through IBD interactions could reach 12 degrees (68\% C.L) for a supernova located \SI{10}{\kilo\parsec} away.

It is worth noting that efforts are underway to extract directional information from the small amount of Cherenkov light which leads the largely isotropic scintillation light~\cite{Aberle:2013jba}.  The CHESS experiment found that a time resolution of $338\pm 12$ ps (FWHM) was required for reasonable efficiency in separating the two light components in a mixture of LAB with 2g/L PPO~\cite{Caravaca:2016ryf}.  An alternative to fast PMT's is to slow the emission of scintillation light (e.g., \cite{Wang:2017etb,Biller:2020uoi}), which is possible with different scintillators and fluors.  Finally, the different spectra of the two components may be exploited using dichroic filters~\cite{Kaptanoglu:2018sus,Kaptanoglu:2019gtg}.  Such ideas may be exploited in future liquid scintillator detectors, or upgrades of current ones.

\subsection{Triangulation}
\label{sec:triangulation}

\begin{figure}[ht]
\centering
\includegraphics[scale=0.35]{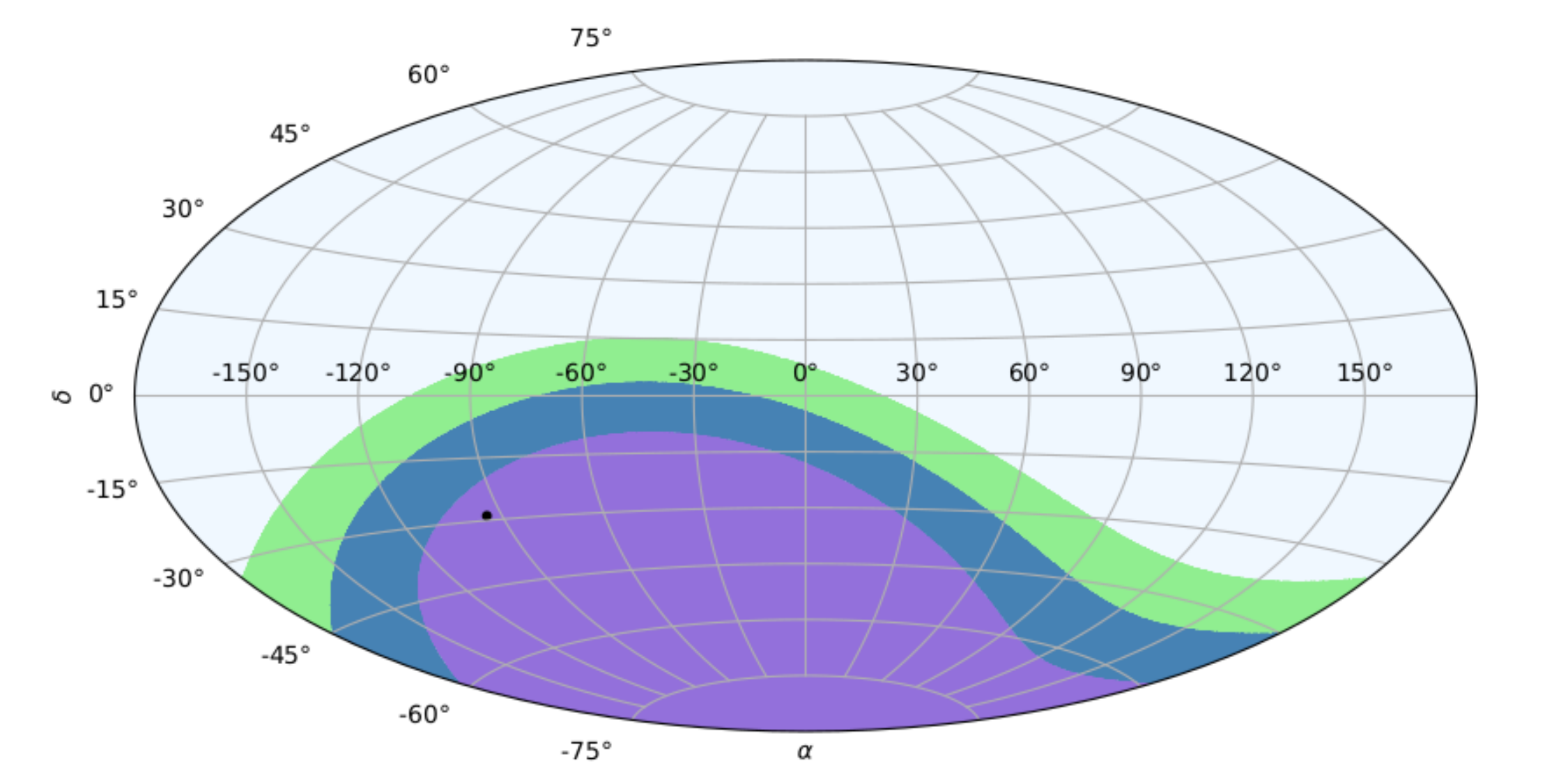}\includegraphics[scale=0.35]{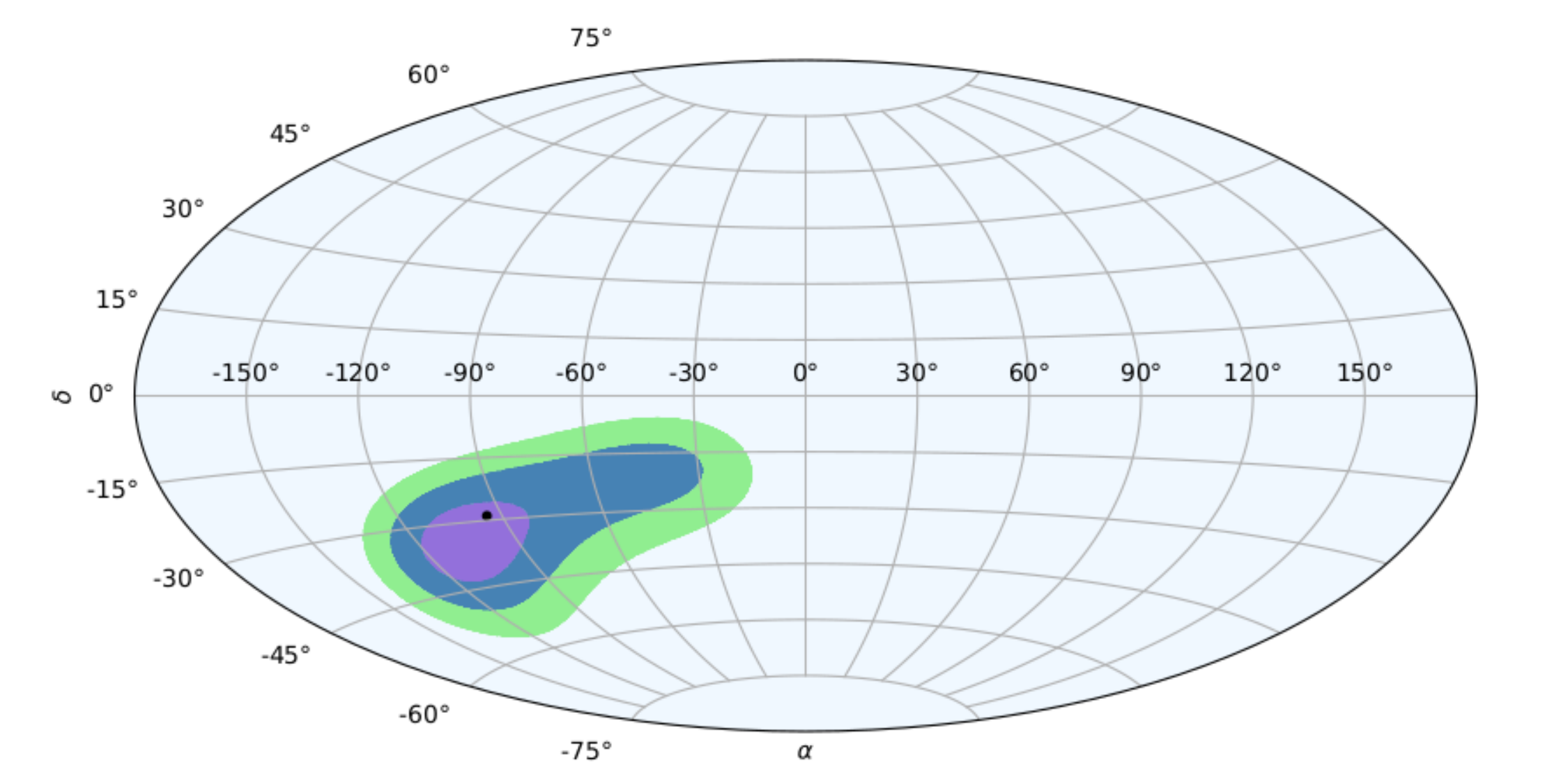}
\caption{Left: Sky area determined at 1$\sigma$ by combining IceCube timing information with Super-K, assuming normal hierarchy and the~\cite{Huedepohl:2014} model for a supernova at \SI{10}{\kilo\parsec}. The true direction is shown with a black dot. Right: Sky area determined by combining IceCube, DUNE, JUNO, and Hyper-K~\cite{Linzer:2019swe}.}
\label{fig:triangulation}
\end{figure}

In triangulation, the time delays of neutrino events observed between detectors at different geographical locations are used to infer the direction of the supernova. For a pair of detectors $i$ and $j$, the delay between them $\Delta t_{ij}$ is defined as follows:
\begin{equation}
\Delta t_{ij} = \vec{d_{ij}} \cdot \vec{n} / c,
\label{eq:deltat_ij}
\end{equation}
where $\vec{d_{ij}}$ is the vector connecting two detector sites and $\vec{n}$ is the unit vector defining the direction of the CCSN. The vector $\vec{n}$ is calculated from the right ascension, $\alpha$, declination, $\delta$, of the source in the geographic horizontal coordinate system, and the event Greenwich mean sidereal time, $\gamma$, expressed as an angle~\cite{Coleiro:2020vyj}:
\begin{equation}
\vec{n}=
(-\cos(\alpha-\gamma)\cos\delta,\,-\sin(\alpha-\gamma)\cos\delta,\,-\sin\delta).
\label{eq:sndir}
\end{equation}
For simplicity, in recent studies \cite{Brdar:2018zds,Linzer:2019swe,Coleiro:2020vyj} 
$\gamma$ has been fixed to $0^{\circ}$; we note that the results are expected to be qualitatively
insensitive on the choice of such parameter, especially for supernovae in the galactic center (where $\alpha$ is large).
For a known $\Delta t_{ij}$ and $\vec{d_{ij}}$, Eq.~\ref{eq:deltat_ij} defines a cone that has a thickness $2 \delta(\cos \theta_{ij})$ due to the uncertainty $\delta (\Delta t_{ij})$. Typically, $\Delta t_{ij} \approx$ \SI{30}{\milli\second} for pairs of neutrino detectors since the Earth diameter corresponds to a time delay of $\sim$\SI{40}{\milli\second}. The uncertainty $\delta (\Delta t_{ij})$ can be evaluated for each detector pair as in \cite{Linzer:2019swe, Coleiro:2020vyj}; or as $\delta( \Delta t_{ij}) = \mathrm{Max}(\delta t_i, \delta t_j)$, where $\delta t_i$, $\delta t_j$ are each detector uncertainties defined independently as in \cite{Brdar:2018zds}. The probability that a test position in the sky ($\alpha$, $\delta$) coincides with the equatorial coordinates of the CCSN can be evaluated with the following $\chi^{2}$ function:
\begin{equation} 
\chi^{2}_{ij}(\alpha,\delta) = 
\left( 
\frac{\Delta t_{ij}(\alpha,\delta)-\Delta t_{ij}^\mathrm{data}}{\delta (\Delta t_{ij})}
\right)^2 \: ,
\label{eq:chi2tri}
\end{equation}
the minimum of the function gives the best estimate for the angles ($\alpha$, $\delta$) for the searched CCSN location in the sky. 

Different detector pairs can be combined into a total $\chi^{2}$ by summing each contribution:
\begin{equation} 
\chi^{2}(\alpha,\delta) = \sum_{i,j}^{i<j} \chi^{2}_{ij}(\alpha,\delta) \: .
\label{eq:chi2tot}
\end{equation}
The $\chi^{2}(\alpha,\delta)$ function is converted into a p-value, $p(\alpha,\delta)=p(\chi^{2}(\alpha,\delta)\le\chi^{2}_\mathrm{min})$, which is the probability of observing a $\chi^{2}$ smaller or equal to the minimum $\chi^{2}$ value obtained scanning all possible directions $(\alpha, \delta)$. The 90\% confidence level (C.L.) error box of the source localization area is determined as a collection of all points on the sky with a p-value $p(\alpha,\delta)<0.9$.

Figure~\ref{fig:triangulation} shows the results of some early studies~\cite{Linzer:2019swe} performed using the time of arrival of the first events of a burst in each detector simulated with SNOwGLoBES, and correcting for biases due to relative event rates in each detector. This method is quite robust against flux uncertainties. This approach requires additional data processing for long-string Cherenkov detectors, such as IceCube and KM3NeT, where an event-by-event reconstruction is not feasible. 

\begin{figure}
\centering
\includegraphics[width=0.47\textwidth]{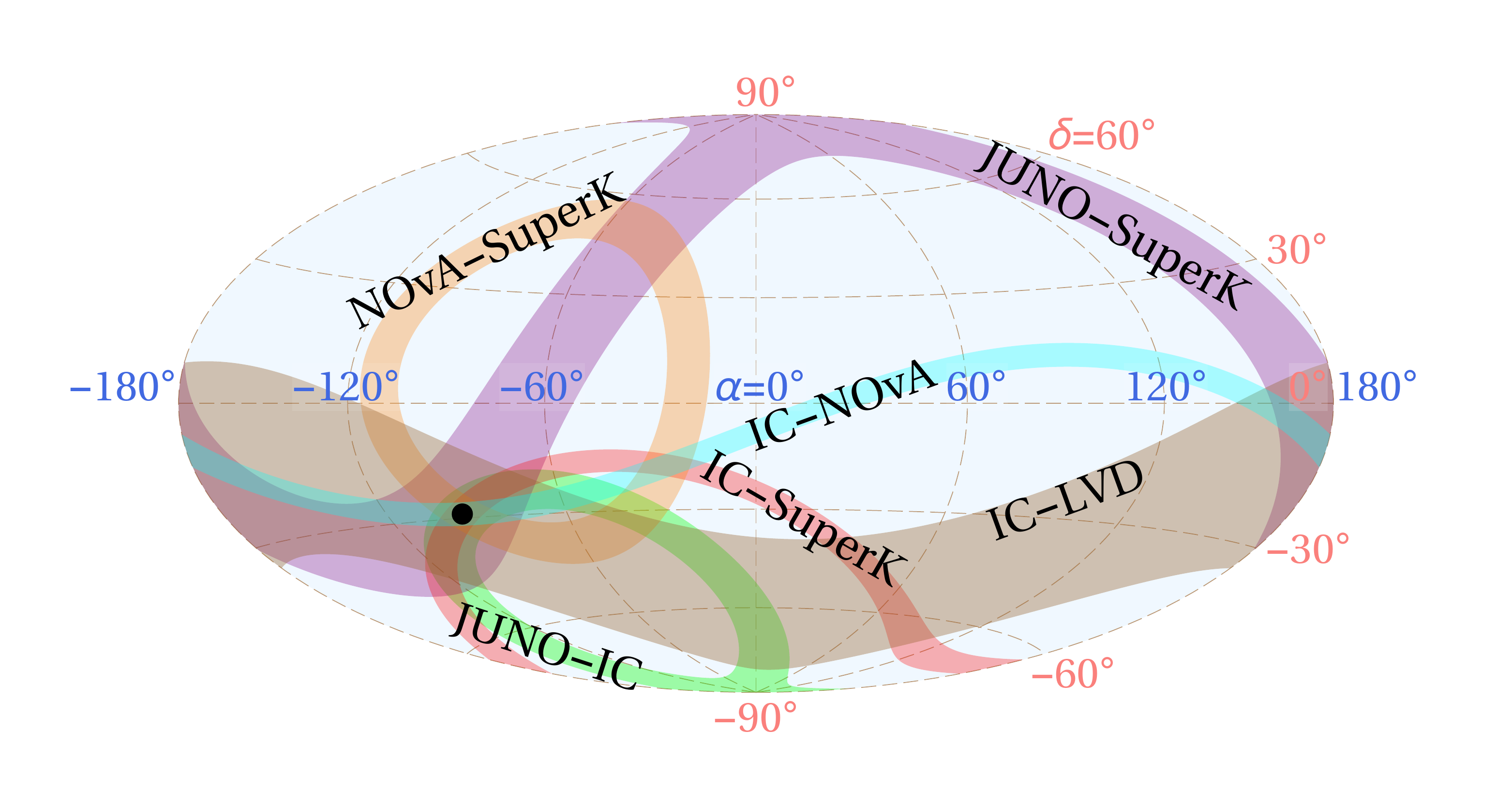}\ \ \includegraphics[width=0.47\textwidth]{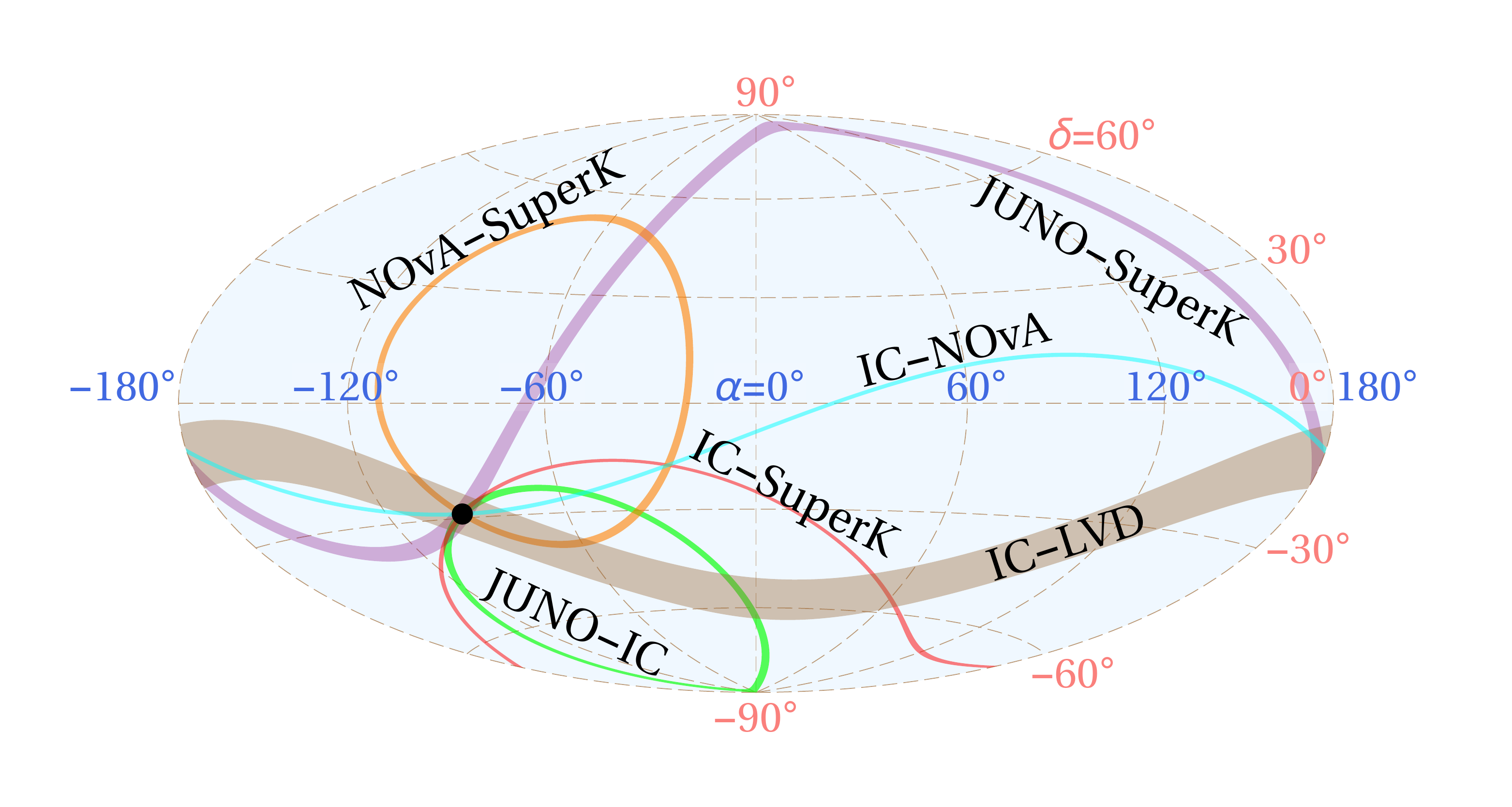}\caption{Regions constrained at $1\sigma$ CL by two-detector combinations, adopted from Ref.~\cite{Brdar:2018zds}.  The left and right panels show scenarios of the supernova core-collapse into a neutron star or a black hole, respectively.  All regions expectedly overlap at the supernova location (black dot) which are set in the Galactic center. \label{fig:brdar_triangulation}}
\end{figure}

\begin{table*}
\centering
\begin{tabular}{ccccccc}
Experiments & major process & target &  $\delta t$  &  $\delta t$ (BH)\\
\hline
Super-K  & $\overline{\nu}_{e}+p\rightarrow e^{+}+n$                 & \SI{32}{\kilo\tonne} ${\rm H}_{2}{\rm O}$       & \SI{0.9}{\milli\second}  & \SI{0.14}{\milli\second} \\
JUNO     & $\overline{\nu}_{e}+p\rightarrow e^{+}+n$                 & \SI{20}{\kilo\tonne} ${\rm C}_{n}{\rm H}_{m}$    & \SI{1.2}{\milli\second}  & \SI{0.19}{\milli\second} \\
DUNE     & $\nu_{e}+{\rm ^{40}Ar}\rightarrow e^{-}+{\rm ^{40}K^{*}}$ & \SI{40}{\kilo\tonne} LAr                         & \SI{1.5}{\milli\second} & \SI{0.18}{\milli\second} \\
NO$\nu$A & $\overline{\nu}_{e}+p\rightarrow e^{+}+n$                 & \SI{14}{\kilo\tonne} ${\rm C}_{n}{\rm H}_{m}$    & \SI{1.4}{\milli\second}  & \SI{0.24}{\milli\second} \\
CJPL     & $\overline{\nu}_{e}+p\rightarrow e^{+}+n$                 & \SI{3}{\kilo\tonne} ${\rm H}_{2}{\rm O}$         & \SI{3.8}{\milli\second} & \SI{0.97}{\milli\second} \\
IceCube  & noise excess                                              & ${\rm H}_{2}{\rm O}$                             & \SI{1}{\milli\second}    & \SI{0.16}{\milli\second} \\
ANTARES  & noise excess                                              & ${\rm H}_{2}{\rm O}$                             & \SI{100}{\milli\second}  & \SI{32}{\milli\second}   \\
Borexino & $\overline{\nu}_{e}+p\rightarrow e^{+}+n$                 & \SI{0.3}{\kilo\tonne} ${\rm C}_{n}{\rm H}_{m}$   & \SI{16}{\milli\second}      & \SI{5.5}{\milli\second}  \\
LVD      & $\overline{\nu}_{e}+p\rightarrow e^{+}+n$                 & \SI{1}{\kilo\tonne} ${\rm C}_{n}{\rm H}_{m}$     & \SI{7.5}{\milli\second}     & \SI{2.4}{\milli\second}  \\
XENON1T  & coherent scattering                                       & \SI{2}{\tonne} ${\rm X_{e}}$                     & \SI{27}{\milli\second}            & \SI{10}{\milli\second}   \\
DARWIN   & coherent scattering                                       & \SI{40}{\tonne} ${\rm X_{e}}$                   & \SI{1.3}{\milli\second}      & \SI{0.7}{\milli\second}  \\
\end{tabular}
\caption{A summary of supernova neutrino arrival time
uncertainties ($\delta t$) estimated in Ref.~\cite{Brdar:2018zds}. In the second and third columns, the main detection channel as well as the target are shown for each of the experiments listed in the first column. The next (last) two columns show the $\delta t$ values for the galactic supernova core-collapse into a neutron star (black hole).}
\label{tab:brdar2018}
\end{table*}

The authors of \cite{Brdar:2018zds} have demonstrated that by fitting the temporal
shape of the neutrino flux, the uncertainty on the supernova onset time, $\delta t$, decreases in comparison to what was reported in earlier studies~\cite{Beacom:1998fj} where, in the absence of results from supernova simulations that we have nowadays, simple forms for the event rates were assumed. See also~\cite{Hansen:2019giq} for a recent timing analysis. 

Fig.~\ref{fig:brdar_triangulation} shows the $1\sigma$ regions of supernova directions constrained by several two-detector combinations. The left and right panel correspond to the  case of core-collapse into a neutron star and black hole, respectively. Table~\ref{tab:brdar2018} summarizes arrival time uncertainties for a number of present and future neutrino detectors, for neutron star and black hole final states. The advantage of this method is most evident in cases with rapid temporal variation, in particular the sharp cut-off in the flux arising from the formation of a black hole. Namely, it was found in \cite{Brdar:2018zds} that a small fraction of events around the cut-off chiefly determines the timing uncertainty in this scenario (while the events around the onset were also considered, their effect turned out to be marginal for obtaining $\delta t$ in the performed statistical analysis). The disadvantage of the applied fit is its dependence on the theoretical prediction of the flux.

Ideally, one would want to exploit advantages of both \emph{first-event} method \cite{Linzer:2019swe} and the $\chi^{2}$ \emph{fit of the full-spectrum} \cite{Brdar:2018zds}. For instance, including the first couple of events in the fit (or last few events in case of the black hole scenario), would be less model-dependent than the latter and statistically more robust than the former.

\begin{figure}[!ht]
    \centering
    \includegraphics[width=0.5\linewidth]{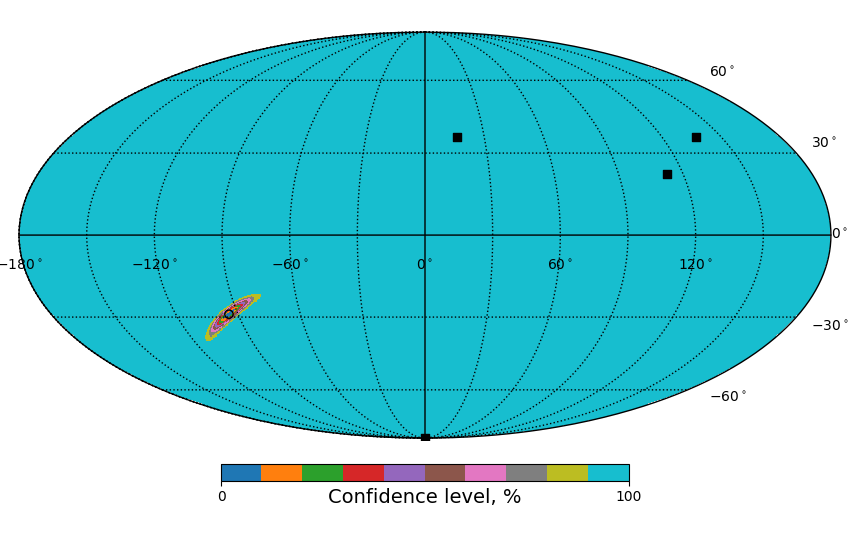}
    \caption{Confidence area in equatorial coordinates for a CCSN at the Galactic Center (black dot) computed using triangulation between four detectors: IceCube, KM3NeT/ARCA, Hyper-Kamiokande and JUNO. Their position is indicated with the black squares. Figure from~\cite{Coleiro:2020vyj}.}
    \label{fig:skymap4detectors}
\end{figure}

In order to reduce model dependency for the $\chi^{2}$ \emph{fit of the full light-curve}, direct matching of the detected neutrino light-curves has been explored in~\cite{Coleiro:2020vyj} using two different techniques to evaluate the signal arrival time and its uncertainty: $\chi^{2}$ and normalized cross-correlation. The results reproduced in Fig.~\ref{fig:skymap4detectors} and Table~\ref{t:results} show that an uncertainty area of $\sim$70~deg$^{2}$ (at 1$\sigma$ level) in the sky can be achieved when combining four current and near-future detectors sensitive to IBD (IceCube, KM3NeT-ARCA, Hyper-Kamiokande and JUNO). Techniques for further data-driven optimization, once the supernova has been observed, have also been investigated. Systematic effects will be explored using detailed core collapse supernova explosion models and more realistic detector descriptions.

\begin{table}[!ht]
\centering
\resizebox{\linewidth}{!}{
 \begin{tabular}{ccccc}\hline   
 & {\bf KM3NeT/ARCA} & {\bf Super-Kamiokande} & {\bf Hyper-Kamiokande} & {\bf JUNO} \\ \hline
{\bf IceCube} & \,6.65$\pm$0.15 & \,1.95$\pm$0.04 & \,0.55$\pm$0.01 & \,1.95$\pm$0.04\\ \hline
{\bf KM3NeT/ARCA} & - & \,7.4$\pm$0.2  & \,6.70$\pm$0.15 & \,7.4$\pm$0.2 \\ \hline
{\bf Super-Kamiokande} & - & - & - & \,2.75$\pm$0.06 \\ \hline
{\bf Hyper-Kamiokande} & - & - & - & \,1.99$\pm$0.04 \\ \hline
  \end{tabular}
   } 
\caption{Uncertainty $\delta t$ in milliseconds obtained with the chi-square method using average background subtraction and unity normalization of the detector neutrino light-curves. The detector pairs are listed in row and column names.}
\label{t:results}
\end{table} 

\section{Presupernova Neutrinos}
\label{sec:preSN}

Several days before core collapse begins, neutrino production in the core and inner shells of the star increases dramatically, as nuclear fusion proceeds through its final stages of carbon, oxygen and silicon burning. These pre-supernova neutrinos are due to enhanced thermal emission~\cite{Odrzywolek:2003vn, Odrzywolek:2004em, Odrzywolek:2010zz} --- as 
the temperature inside the star increases progressively --- and to beta processes involving a large network of nuclear species \cite{Patton:2015sqt,Kato:2017ehj,Patton:2017neq}.  The emissivity is dominated by $\nu_e$ and $\bar \nu_e$ (the flux of non-electron neutrino species will be comparable after flavor conversion inside the star); their energy spectra are typically sub-MeV, with a peak at $\sim$ \SIrange{1}{3}{\mega\electronvolt} in the last hours of the emission. 

\begin{figure}[ht]
  \centering
  \includegraphics[width=0.8\textwidth]{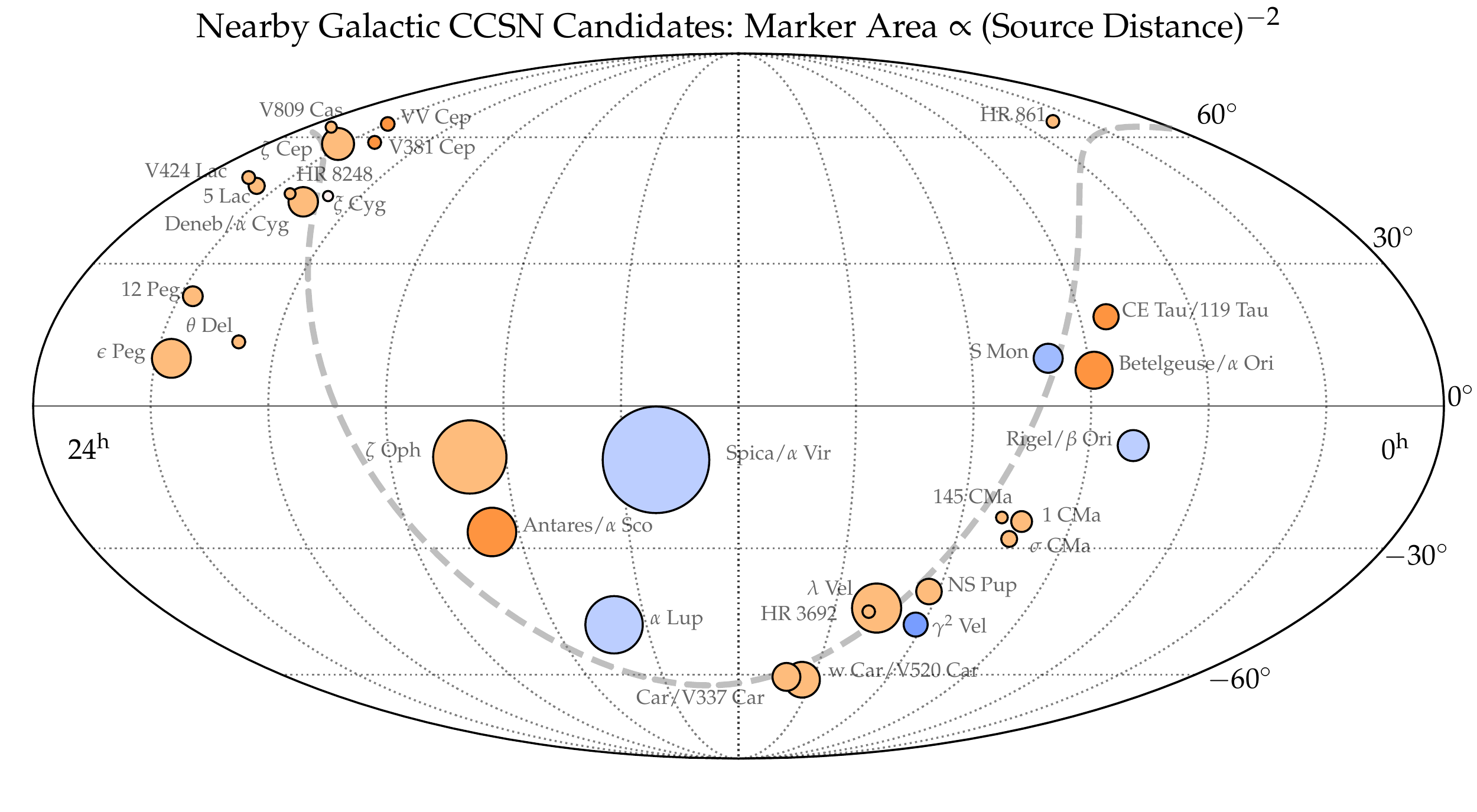}
  \caption{Equatorial coordinates of supernova-ready blue and red supergiant stars within $\sim$\SI{1}{\kilo\parsec} of Earth (adapted from \cite{Mukhopadhyay:2020ubs}).}
  \label{fig:presupernova_candidates}
\end{figure}

The total energy emitted in pre-supernova neutrinos is weak, orders of magnitude below the energy of the post-collapse burst. Therefore, its detectability is limited to nearby stars, at distances up to $\sim$\SI{1}{\kilo\parsec} (see, {\it e.g.},~\cite{Yoshida:2016imf,Patton:2017neq}). There are about 40 supernova-ready stars (stars that are already in the supergiant stage) within this radius from Earth, the best known of which are Betelgeuse and Antares.  The detection of pre-supernova neutrinos will thus be a much rarer event than the detection of a supernova burst; considering its exceptional character, every possible measure should be put in place the ensure it is not missed. A pre-supernova alert, made hours before the collapse,  would provide precious extra time to prepare multi-messenger observations of the collapse and ensure all detectors are taking data (in neutrinos and gravitational waves, and possibly exotic particles, like the axion) and of the explosion of the star. It would be especially important if the collapsing star has already shed its envelope and is left with a relatively low mass, resulting in a short time (possibly under one hour, see, {\it e.g.},~\cite{Muller:2018utr}) between the collapse and the explosion. 

A methodology for a pre-supernova alert has been implemented in KamLAND~\cite{Asakura:2015bga}. It is estimated that it could provide a 3$\sigma$ detection 48 hours prior to the explosion of a \SI{25}{\solarmass} star at \SI{690}{\parsec}. Extending this alert to those experiments in the network with low enough thresholds should expand the sensitivity to a larger fraction of the galaxy.  At this time, the potential to identify the progenitor star --- possibly using a combination of directional neutrino detections and theoretical priors --- has not yet been studied.
In the near future, Super-Kamiokande will be loaded with gadolinium to improve its neutron tagging efficiency. Once completed, its sensitivity to pre-supernova neutrinos is expected to be similar to that of KamLAND~\cite{Simpson:2019xwo}.

Large direct dark matter detection experiments based on argon and xenon can constitute efficient pre-supernova detectors~\cite{Raj:2019wpy}, because of scalable fiducial volumes as well as very low  thresholds. Due to heavy nuclear targets, these experiments can take full advantage of coherent neutrino-nucleus scattering from pre-supernova neutrinos, allowing to bypass the kinematic threshold limiting IBD scattering to neutrinos with energies above \SI{1.8}{\mega\electronvolt} as well as sensitivity to all neutrino flavors. Hence, large dark matter detectors can provide complementary pre-supernova neutrino information to that of dedicated neutrino experiments.

\section{The SNEWS Alert and Followup}
\label{sec:snews}

The science of SNEWS depends upon developing and sustaining a software stack with a number of separate but interrelated components: 
\begin{itemize}
    \item A neutrino  data aggregator
    \item A platform for analyzing this data to generate alerts
    \item A system for tracking and updating alerts from SNEWS~2.0 and member experiments
    \item A system for combining and summarizing alerts intelligently
    \item A system for distributing and archiving alerts
\end{itemize}

\textbf{Aggregating Neutrino Data:} SNEWS~1.0 has a long track record of receiving certified, confidential burst alerts from neutrino detectors. These alerts currently contain very little information about the burst, but it should be straight-forward to expand the types of information that can be reported.  For example, neutrino ``light curves'' for triangulation calculations, neutrino-electron scattering directional error boxes, and constant time-series of significances to allow detection of sub-threshold bursts or pre-SN neutrinos.

\textbf{Analyzing Burst Data:} Because SNEWS receives certified data from a large number of experiments, it can serve as a platform for performing analyses that are enhanced by combining information from multiple detectors. The data of course belong to those experiments not to SNEWS, and it is vital that the experiments are happy with the way their data are being used and credited.  Any specific combination analysis could be considered a ``virtual experiment''.  Before an analysis is implemented, each detector participating in a given virtual experiment will have signed MOUs with SNEWS and the other detectors involved in that analysis, so that exactly who is sharing what and what the output will look like is carefully defined in advance.  Some examples of these virtual experiments would be the SNEWS burst trigger itself, a pre-supernova neutrino trigger, a triangulation-based pointing algorithm, and combining skymaps from multiple pointing methods. 

\textbf{Tracking and Updating Alerts:} 
As new information comes in, either updates and refinements from a previously reporting experiment or late-arriving measurements from a different experiment, the alert will be updated.

\textbf{Combining and Summarizing Alerts:} In the event of a galactic supernova, there will be many observation reports coming in very quickly on a number of alert networks. In this case, the goal for SNEWS~2.0 should be to quickly summarize the information from the neutrino community so that it is digestible for the astronomy community. In particular, SNEWS should have a mechanism for intelligently combining sky maps from multiple experiments without human intervention. 


\subsection{Real-Time Algorithmics}
\label{sec:realalgo}

A new suite of software beyond the existing SNEWS coincidence server will need to be designed, written, tested, commissioned, and sustained. 
The success of SNEWS depends on having robust and reliable cyberinfrastructure, which requires developing and supporting intracollaboration code from SNEWS itself while leveraging other prior MMA investments when applicable.
For instance, the  Scalable Cyberinfrastructure to support Multi-Messenger Astrophysics (SCIMMA) project (\url{https://scimma.org/})~\cite{Chang:2019edd} is already developing tools for general purpose multi messenger astronomy software infrastructure, including user management and messaging backend that could be leveraged for SNEWS~2.0.
SNEWS and SCIMMA have created a joint prototype, replicating SNEWS~1.0 functionality using SCIMMA's ``Hopskotch'' toolkit. Other software infrastructure also exist that SNEWS can integrate into, especially to provide machine-readable output to facilite follow-up observations; {\it e.g.}, the Astrophysical Multimessenger Observatory Network (AMON; \cite{Smith13}).

To ensure the provenance of message origin, SNEWS incorporates software infrastructure to facilitate digital certificate generation, signing, distribution, and revocation for the purpose of signing and encrypting neutrino event messages. Messages not signed and encrypted with a valid client certificate are ignored and discarded. This software will need to be upgraded for the enhanced trigger.  Effort needs to be made to ensure reliable time synchronization throughout the participating instruments and analysis hosts. Monitoring and alerting to system time deviations as well as host, network, and detector down times will be developed. 

The triangulation calculations will require access to adequate computation resources in order to process incoming neutrino event messages to generate alerts in a timely manner. Investigation will need to be undertaken to determine the adequacy of existing compute infrastructure for this purpose. Should it prove inadequate, software systems will be developed to leverage additional, existing compute resources for this purpose. Standard software libraries and programming practices will be utilized to the fullest extent possible.

Finally, given the rarity of the occurrence of events this system is designed to detect, end-to-end online testing will need to be integrated into the software design. It should be possible to fully test and verify the system without affecting the live monitoring for neutrino event messages.

\subsection{Multimessenger Follow-Up}
\label{sec:mma-followup}

The part of the long-standing SNEWS project which has always been about true multi-messenger astrophysics has been exploiting the $\sim$hours head start neutrinos have on electromagnetic radiation, to provide astronomers across the electromagnetic spectrum with an early warning, so they can make the best use of the once-in-a-career event of a galactic supernova. The existing simple coincidence of experimental neutrino triggers has no directional information. Currently, only Super-K has any directional sensitivity to $[10-100]\1{MeV}$ neutrinos, via the forward nature of neutrino-electron scattering. This analysis can be published via SNEWS, but not automatically. In the SNEWS~2.0 coincidence network, an automated analysis of the fine timing signals in various detectors has the possibility of producing intersecting error bands on the sky. This can provide direction for astronomers to start looking and thus enhances the prospect that very early light from a supernova, just as the shock breaks out through the photosphere, can be recorded in multiple wavelengths.

Notably, modern transient surveys are now capable of promptly mapping large regions of sky on rapid timescales. In the case of GW170817, an extensive observing campaign of facilities covering the electromagnetic spectrum was able to discover optical counterpart to the merger within 11~hours using 31~deg$^2$ localization provided by the Advanced LIGO and Advanced Virgo detectors. The Zwicky Transient Facility (ZTF), operating in the northern hemisphere, uses custom-built wide-field camera on the Palomar 48 inch Schmidt that provides a 47~deg$^2$ field of view capable of mapping the entire visible sky in $\approx 4$ hours to a limiting magnitude of $r \sim 21$\,mag \cite{Bellm19}. Soon, the Rubin Observatory operating the Legacy Survey of Space and Time (LSST) will be operational. The LSST camera will have a field of view of 9.6~deg$^2$. Though smaller than ZTF, LSST will have substantially improved sensitivity, reaching $ r \approx 24$\,mag in a single 10 second exposure. Wide-field imaging at near-infrared wavelengths is also now possible ({\it e.g.}, Palomar Gattini-IR; \cite{Moore2016}). These individual facilities, along with networks of telescopes already familiar with how to effectively coordinate multi-messenger follow up ({\it e.g.}, Global Relay of Observatories Watching Transients Happen~\cite{Growth2020}; Global Rapid Advanced Network Devoted to the Multi-messenger Addicts~\cite{Grandma2020}) can provide precise localization at optical and near-infrared wavelengths that other ground- and space-based observatories sensitive to emission from the supernova across the electromagnetic spectrum can act upon.

Additionally, gravitational wave astronomy is now a reality. Gravitational waves also escape the nascent supernova promptly. However, unlike merging compact objects, the expected signal of a core collapse supernova in gravitational waves is highly dependent upon asymmetries in the matter distribution. A symmetric collapse, even if nearby, could make very little signal in gravitational wave detectors, but whatever signal it does provide will be an important component to understanding the supernova itself. Knowing the detailed neutrino ``light curve'' as soon as possible will help the gravitational wave community to unravel what they are seeing in their detectors, facilitating rapid followup campaigns.   
In particular, by using the same analysis used to do triangulation, SNEWS~2.0 can provide, to the GW community, the temporal window in which to look for the GW signal. The gain that this information produces has been investigated in~\cite{Nakamura:2016kkl} for a CCSN located in the Galactic center and emitting the GW signals of Fig.~\ref{fig:multimessengersignal}. The observation of the CCSN with neutrinos from Super-K alone allows the identification of a time window of \SI{60}{\milli\second} around the time of the bounce where the GW signal is expected; the SNEWS~2.0 triangulation goal (Sec.~\ref{sec:triangulation}) hopes to identify this time to an order of magnitude tighter precision. By using this information the SNR of the GW signal increases from $\sim 3.5$ to $\sim 7.5$ (for  \SI{60}{\milli\second} precision), expanding the gravitational detection horizon.  This is especially important for GW from CCSN, since the amplitude of the GW from CCSN are a strong function of the asymmetry in the collapse, and so could be weak even for a galactic SN.

With the detection of gravitational waves, and a better understanding of the mechanism of collapse of the supernova, it could thus be possible to measure the absolute mass of the neutrino via the time difference of detection between the two signals ({\it e.g.}~\cite{Vissani:2010fg}).

\subsection{Alert Broadcasting and Optimized Observing Strategies}
\label{sec:alert-broadcasting}

The existing SNEWS project relies on a mailing list of interested individuals and direct connections to experiments (such as NOvA and XENON1T) and projects (GCN) to promulgate any alert to the wider community. SNEWS2.0 will take advantage of new infrastructure for rapid dissemination (see Section~\ref{sec:realalgo}).

Established communication networks (Astronomer's Telegram, LIGO-Virgo Collaboration Alerts) will also be part of the dissemination network. Broadcasting of alerts with the SNEWS~2.0 mobile app will reach both professional and public audiences.

Alerts under SNEWS~2.0 will be accompanied by suggestions for optimized observing strategies and alerts to suitable facilities. Given the wide range of phenomena that may trigger an alert and the short turn-around time for response, pre-planned strategies for coordinated response of facilities are necessary to avoid missed scientific opportunities that may result from observers acting independently. Detection, location, and information regarding explosion type ({\it e.g.}, formation of neutron star vs.\ black hole) will inform follow-up strategies and ensure that it is optimized to maximize scientific return. Strategy suggestions will leverage the Recommender Engine for Intelligent Transient Tracking (REFITT) being developed at Purdue University \cite{Sravan20}. REFITT is an Artificial Intelligence transient inference and strategy engine that designs and co-ordinates optimal follow-up of supernova events in real-time. Having observing strategies that leverage an observing alliance of professional and amateur observers that join the SNEWS~2.0 response network will maximize use of available technological resources while reducing redundancy and missed observing opportunities, enabling the extraction of as much science as possible, particularly during the first few hours following core collapse. 

\subsection{Latency}
\begin{figure}
    \centering
    \includegraphics[width=0.7\columnwidth]{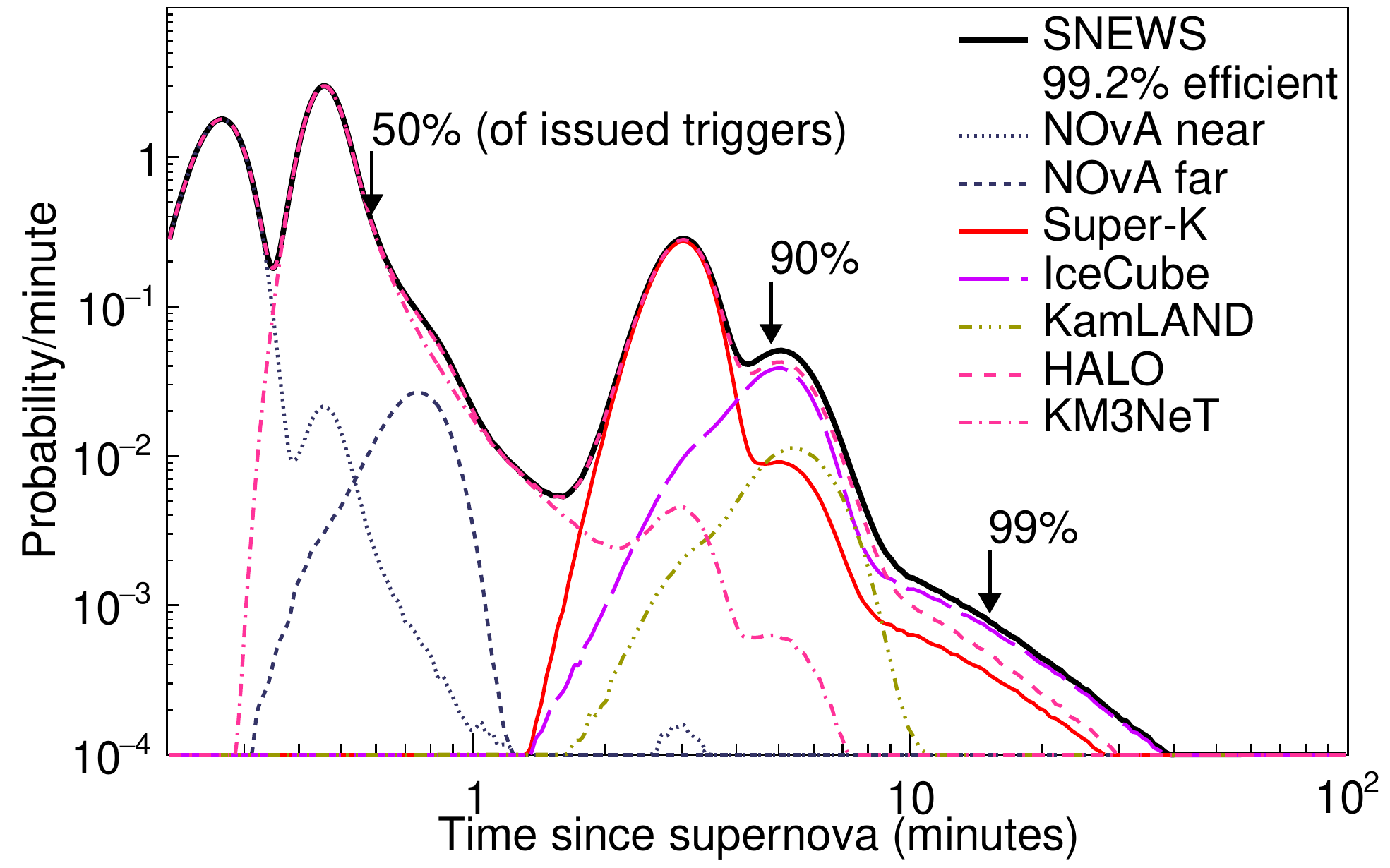}
    \caption{Estimated latency (horizontal axis) for an example configuration of SNEWS~1.0 including the set of experiments most likely to be in the network in 2021.  Each experimental curve is the probability of an alert from that experiment arriving at SNEWS at the given time and being one of the first two experiments to do so~\cite{novaSN,Abe:2016waf,Agafonova:2007hn,Wei:2015qga}  The combined curve is the probability of a SNEWS alert being formed. }
    \label{fig:latency}
\end{figure}

Another challenge concerns timing latency. While the delay time between the neutrino emission and the first electromagnetic signal (arising from the shock breakout) is typically considered to be on the scale of hours, as it was in SN1987A, this is only true for the supernova explosion of supergiant stars such as the one shown in Fig.~\ref{fig:multimessengersignal}. However, not all massive stars end their lives as supergiants, and the delay time may be much less. The delay time has dependencies on multiple parameters including the stellar interior structure and explosion energy, but is largely driven by the stellar envelope radius which the shock wave must cross before it can burst out of the star's photosphere. 

In the local Universe, some 25\% of core-collapse supernovae are Types Ib or Ic \cite{Li:2010kc}, indicating that the progenitor stars have shed their hydrogen and/or helium envelopes prior to explosion. These stars, known as Wolf-Rayet stars, have radii that are some $\sim 1/100$ compared to supergiants, thus reducing the time delay between neutrino emission and first light to a mere minutes. This places stringent requirements on real-time analysis, on the release of the neutrino trigger information, and on established follow-up strategies. The inputs to the present SNEWS network will need to be upgraded in order to meet these stringent requirements, which will require dedicated inter-experiment coordination and collaboration.

As an example of the risk inherent in the current system, Fig.~\ref{fig:latency} shows an estimate of the probability as a function of time that a SNEWS alert is issued for a set of participating experiments similar to that expected in late 2020 or 2021: NOvA, Super-Kamiokande, IceCube, HALO, KamLAND and KM3NeT.  Estimated livetimes and galactic coverage for each participant have been used to produce this estimate, along with available information for latency, or reasonable estimates when this is not available.  Each experiment's curve in this figure is the probability of it being one of the first two to send its trigger to SNEWS, since the latency of the current SNEWS system is driven by how long it takes two experiments to report a supernova, forming a coincidence.  The calculation of these probabilities is done by a toy Monte Carlo, drawing supernova distances from the distribution of progenitors, combining this with published experimental sensitivity curves and livetime fractions, and then drawing a latency from a distribution corresponding to published experimental trigger reaction times.  NOvA, KM3NeT and HALO typically send their triggers to SNEWS in under a minute.  If two of these experiments are live and the supernova is near enough (all three of these fast experiments also have a limited range), a very rapid alert can be issued.  Otherwise, the alert is much slower and the probability of sending it does not reach 99\% until nearly 20 minutes after the neutrino burst.  This delay is partly a consequence of the SNEWS~1.0 requirement of once-per-century false alarm rates, which imposes a high burden on individual experiments to vet their data, sometimes involving human intervention.  Tolerance of a higher false alarm rate is therefore essential to delivering timely alerts.

\subsection{Data Sharing}
\label{sec:dataSharing}

The extent to which experiments share data will be key to how much we can learn from the next galactic core-collapse supernova. In the current version of SNEWS, experiments send a packet stating that they have seen something consistent with a burst of supernova neutrinos at some time.  If multiple experiments see a burst within \SI{10}{s}, an alert will be issued (Sec.~\ref{sec:current_config}).  More information exchanged would make the most of this rare opportunity. For example, individual experiments' neutrino ``light curves'' may provide the quickest, albeit rough, triangulation (Sec.~\ref{sec:triangulation}) which can help prepare telescopes set up for electromagnetic observations.  Features evident only in the combined light curves may also influence follow-up strategy (see Sec.~\ref{sec:alert-broadcasting}), and a precise neutrino arrival time can help define a gravitational wave search in the case of spherical symmetry producing smaller waves (Sec.~\ref{sec:mma-followup}).

Extensive data sharing may be even more important for the detection of low-flux events such as SNIa, pair instability SNe, and pre-supernova neutrinos, which may yield only marginally significant signals in any individual experiment. If experiments share rolling updates of their ``pre-supernova significance'', their combination would increase the sensitive range and advance warning time for the explosion, and indeed provide important information on the dynamics within the supernova progenitor (see Sec.~\ref{sec:lower_threshold}).

Participating experiments will choose the degree of data sharing that they are most comfortable with.  Functional levels of data sharing for a given detector can be divided into three example tiers:
\begin{enumerate}
\item {\bf Alert Tier:} The detector sends a message to SNEWS~2.0 indicating that it sees above-threshold activity. Detectors could also send status messages to indicate whether they are taking data or not; the status messages are used to evaluate the joint significance of coincident activity, and can also be used to avoid collective down-times.
\item {\bf Significance Tier:} In addition to sending Alert Tier messages, the detector sends messages indicating the signal significance and other aggregate characteristics of current observations (or null observations), {\it e.g.}, a $p$-value for background or signal (such as the ``pre-supernova significance''), or a skymap of $p$-values when relevant. These messages can be sent periodically or when the significance changes rapidly (albeit below the threshold of the Alert Tier).
\item {\bf Timing Tier:} In addition to Alert and Significance Tier messages, the detector sends information related to the time series underlying the Significance Tier information. The time series may consist of individual event information (such as time stamps and energy), or a distribution of events binned in time, as appropriate for the detector. Different interaction channels are sent in different time series.
\end{enumerate}

Individual detectors may opt to share at different tiers for pre-supernova and supernova data.  In the case of pre-supernova data, it is clear that Significance Tier sharing allows SNEWS~2.0 to extend
sensitivity beyond what each individual experiment can manage, while Timing Tier sharing could provide further information on what may be expected from the subsequent burst.

For supernova data, the Alert Tier is similar to what is already done in the original SNEWS.  Significance Tier sharing enables pointing back to the supernova, since it is at this level detectors might share derived quantities such as a common $t_0$ or a pointing based on anisotropic interactions. The drawback is that such derivations may take considerable time, introducing a latency which could limit its usefulness. Timing Tier sharing, on the other hand, is intended to gather information which is available with low latency and from which a rough pointing can be derived; this could help telescopes prepare their observation strategies, which can then be refined with further data (some of which may fall under Significance Tier sharing).  As mentioned earlier, Timing Tier data sharing may also reveal features which would be important in multi-messenger follow-up.

All the data shared with SNEWS remains the property of the originating experiment, and each experiment will have its own connection with the SNEWS server which will be private from all other users. SNEWS will store
different categories of shared data for the following periods of time: one month, for all Alert Tier data, and overall $p$-values under the Significance Tier, in order to catch irregularities; 48 hours, for pre-supernova data under Significance and Timing Tiers; and one hour, for supernova data under Significance and Timing Tiers.

\begin{figure}
    \centering
    \includegraphics[width=0.9\columnwidth]{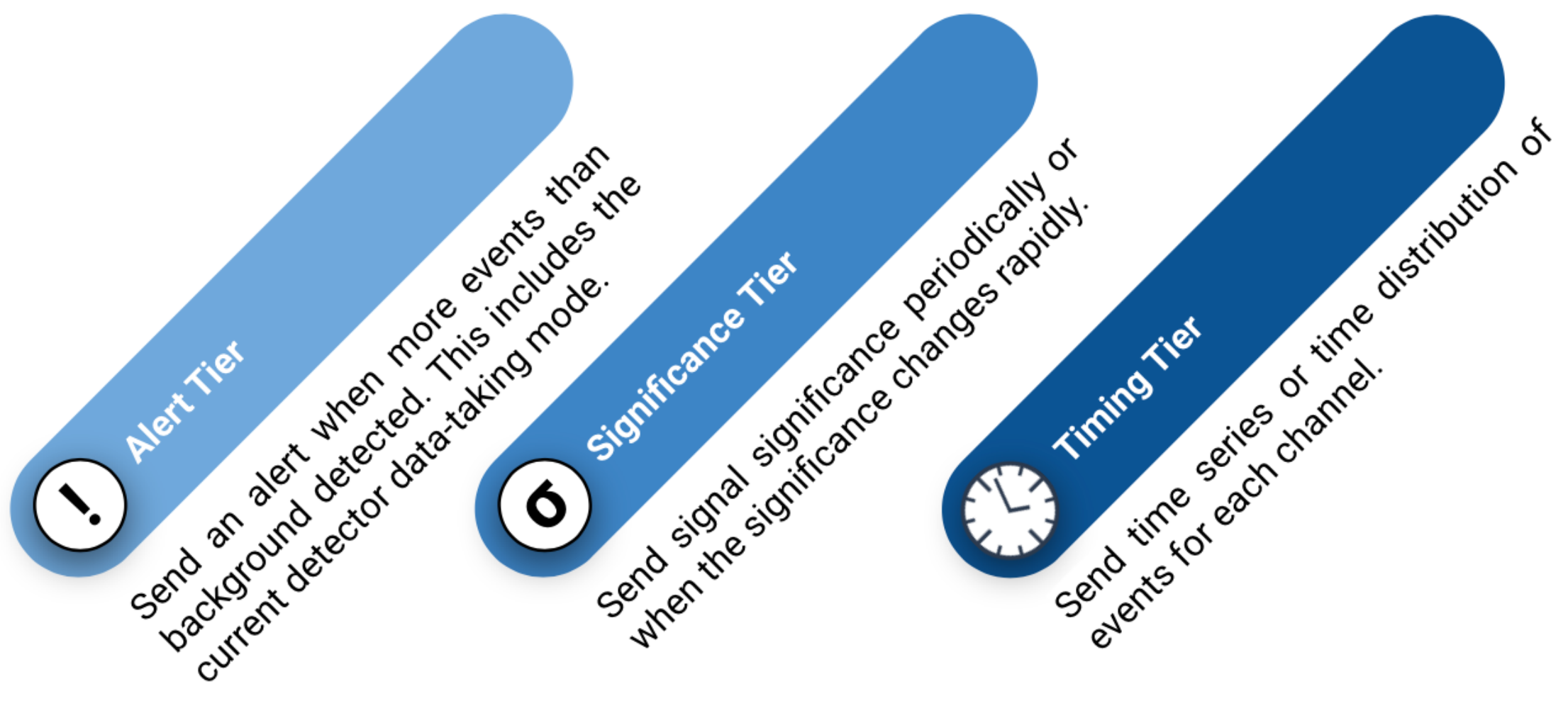}
    \caption{The three example tiers of data sharing, increasing in complexity (and usefulness) from left to right, as described in Sec.~\ref{sec:dataSharing}.}
    \label{fig:detector_contributions}
\end{figure}

\subsection{Walkthrough Example}

\begin{figure}
    \centering
    \includegraphics[width=0.7\columnwidth]{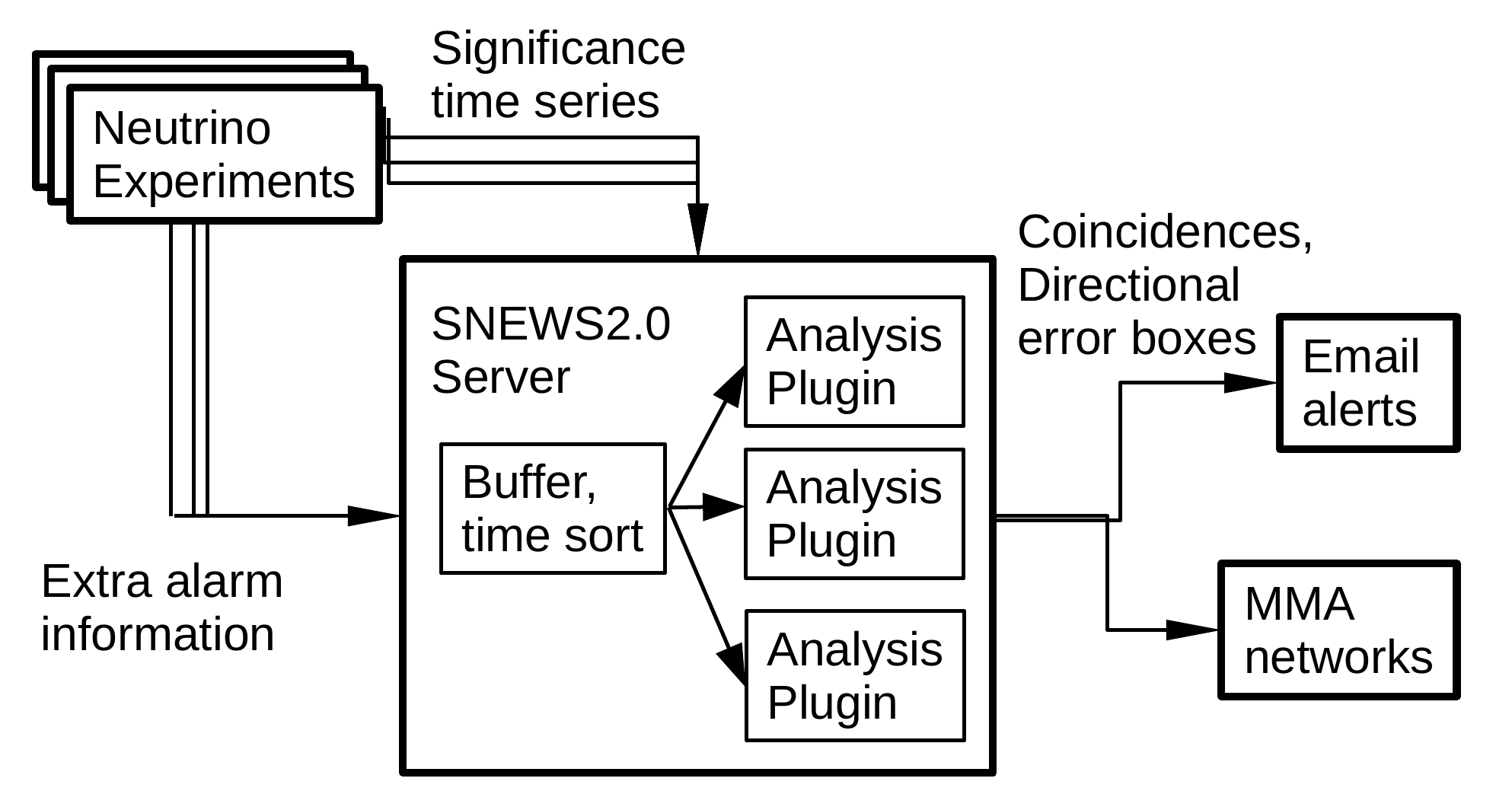}
    \caption{An example flowchart of how the new SNEWS~2.0 network might function.  Each neutrino experiment will contribute to one or more streams of information.  The server will collate, buffer, and sort that information, which analysis plugins will operate on to produce various outputs.}
    \label{fig:blocks}
\end{figure}

While the details of the implementation are still being developed, a block diagram of the flow of an improved SNEWS~2.0 network is shown in Fig.~\ref{fig:blocks}.  Each experiment can contribute to one or more data streams, {\it e.g.}, a time series of the significance of a supernova neutrino signal over time in that experiment, or a traditional alert to be used to form a coincidence, as in the original SNEWS.  In the event of an alert, extra information to be used in triangulation could be sent, or neutrino ``lightcurves", distance estimates, or other useful items.

The server itself will sort, collate, and buffer the incoming information, making it available to analysis plugins that operate on that data.  One plugin will combine significance time series to search for signals that might be sub-threshold in any given experiment.  Another might combine t0's to calculate direction from triangulation, or take and combine maps of error boxes on the sky from experiments capable of finding direction in their neutrino signals.

Outputs will also be varied.  The traditional SNEWS email alert list will be maintained.  Multiple thresholds of alarms will be available.  Also, machine readable outputs formatted for use by existing multi-messenger astronomy networks will be provided, to allow automated followups of any signals detected in neutrinos.
\section{Supernova-Neutrino Sensitive Detectors}
\label{sec:detectors}

There are a growing number of detectors sensitive to neutrinos from a galactic supernova. They vary widely in size, detection strategy, and sensitivity. This section provides an overview of the capabilities of experiments that might contribute to the future efforts described above. This whitepaper simply describes possibilities, the details of any given experiment's participation in SNEWS~2.0 would be defined at a later date in a formal MOU.

\subsection{Water Cherenkov Detectors}

Water Cherenkov detectors are based in the instrumentation of a mass of water with photomultiplier tubes. The principle is exploited in two different detector designs. The first, oriented to low-energy neutrinos, is based on water tanks located in an underground laboratory. This is the case for the current Super-Kamiokande and the future Hyper-Kamiokande experiment. The second makes use of long strings (hundreds of metres) of optical modules to instrument a natural environment, like the Antarctic ice (IceCube), a deep sea site (KM3NeT) or a lake (Baikal). These detectors are primarily aimed to high energy atmospheric and astrophysical neutrinos and can reach the cubic kilometre scale.

Water Cherenkov detectors are mainly sensitive to electron antineutrinos interacting via inverse beta decay with the protons in water. Secondary channels are elastic scattering on electrons and interactions with oxygen nuclei.

\subsubsection{Super-Kamiokande}

The Super-Kamiokande detector consists in a cylindrical stainless-steel tank of \SI{39}{\meter} diameter and \SI{42}{\meter} height, filled with \SI{50}{\kilo\tonne} of pure water~\cite{Fukuda:2002uc}. It is optically separated into an inner detector instrumented with 11,146 Hamamatsu R3600 \SI{50}{\centi\meter} diameter hemispherical PMTs (facing inward and fixed at approximately \SI{2.5}{\meter} from the wall) and an outer detector instrumented with 1,885 Hamamatsu R1408 \SI{20}{\centi\meter} diameter hemispherical PMTs (facing outward and fixed at \SI{2}{\meter} from the wall). For supernova detection, only events in the inner detector are used, corresponding to a mass of \SI{32}{\kilo\tonne}.

The detector is located in the Mozumi mine in the Gifu prefecture, Japan. It lies under Mt. Ikeno (Ikenoyama) with a total of \SI{2,700}{\mwe} mean overburden. Data taking began in April, 1996 and has been separated (at the time of the writing) into six periods from SK-I to SK-V, plus SK-Gd starting in July, 2020. SK-Gd is the first period where gadolinium sulfate is dissolved in the tank (0.02\% at first). This allows the detection of a delayed signal from neutron capture on gadolinium following inverse beta decays.

Supernova neutrinos are mainly detected through the inverse beta decay channel ($\bar{\nu}_e + p \rightarrow e^+ + n$, $\sim 4-8K$ events at \SI{10}{\kilo\parsec}) but elastic scatterings on electrons ($\nu + e^- \rightarrow \nu + e^-$, $\bar{\nu} + e^- \rightarrow \bar{\nu} + e^-$) and interactions on oxygen ($\nu_e + ^{16}O \rightarrow e^- + X$, $\bar{\nu}_e + ^{16}O \rightarrow e^+ + X$) are also detected (with respectively $\sim 120-250$ and $\sim 80-250$ events at \SI{10}{\kilo\parsec}).

The current real-time supernova system only uses events in the SK fiducial volume (inner detector volume reduced by 2 meters from the walls, leading to a fiducial volume of \SI{22.5}{\kilo\tonne}) and with visible energy greater than \SI{7}{\mega\electronvolt}\footnote{This effectively reduces the number of expected events by $\sim 30\%$ with respect to the previous paragraph.}. For each selected event, a \SI{20}{\second} time-window is opened backward in time. The total number of selected events $N_{\rm cluster}$ in this time window is computed, as well as the variable $D$ that identifies the dimension of the vertex position distribution ($D \in \{0,1,2,3\}$, corresponding respectively to point-, line-, plane- and volume-like distributions) \cite{Abe:2016waf}.

If $N_{\rm cluster} \geq 60$ events and $D=3$, a ``golden'' warning is generated, without needing any combination with other detectors before worldwide announcement. If $25 \leq N_{\rm cluster} < 60$ events and $D=3$, a ``normal'' warning is generated and is sent to Super-Kamiokande experts as well as to SNEWS for combination. ``Golden'' and ``normal'' warning correspond respectively to $100\%$ supernova detection efficiency at the Large Magellanic Cloud (LMC) and at the Small Magellanic Cloud (SMC). The direction of the supernova event can be recovered using the elastic scatterings with a typical angular resolution of \SIrange{3}{5}{\degree} (depending on supernova model) for an event at \SI{10}{\kilo\parsec} \cite{Abe:2016waf} and is announced independently by the collaboration in a later notice.

\subsubsection{Hyper-Kamiokande}

The Hyper-Kamiokande detector will be located \SI{8}{km} south of Super-Kamiokande, under the peak of Mt. Nijuugo with an overburden of \SI{1,750}{\mwe}. It will be conceptually very similar to the Super-Kamiokande detector, with an enlarged volume of \SI{260}{\kilo\tonne} (\SI{217}{\kilo\tonne} fiducial for supernova), instrumented with $\sim 40,000$ inward-facing \SI{50}{\centi\meter} diameter PMTs and $\sim 6,700$ outward-facing \SI{20}{\centi\meter} diameter PMTs \cite{Abe:2018uyc}.

The \SI{50}{\centi\meter} PMTs for the inner detector are the newly developed Hamamatsu R12860 model, which improves the timing resolution and detection efficiency by a factor of $2$ compared to the model used in Super-Kamiokande. Alternative designs are currently under consideration, which use $\sim 20,000$ \SI{50}{\centi\meter} diameter PMTs, complemented with several thousand multi-PMT modules, each consisting of multiple \SI{7.7}{\centi\meter} diameter PMTs. An alternative design for the outer detector, which uses a larger number of \SI{7.7}{\centi\meter} diameter PMTs to increase the granularity at reduced cost, is also under development~\cite{Zsoldos:2020nbh}.

For a supernova located at \SI{10}{\kilo\parsec}, $\mathcal{O}$(10-100k) inverse beta decays and $\mathcal{O}$(1k) electron scatterings are expected in the detector, allowing precise determination of the energy and time profile of the event, as well as a pointing accuracy of about \SIrange{1}{1.5}{\degree} \cite{Abe:2018uyc}. By using the reconstructed time and energy of all events, Hyper-Kamiokande will be able to distinguish clearly between different supernova models even at distances of up to \SIrange{60}{100}{\kilo\parsec}~\cite{Migenda:2020rot}.

A special trigger for supernovae is currently being designed for the Hyper-Kamiokande DAQ system and the experiment is expected to start data taking in 2027.

\subsubsection{IceCube}

The IceCube Neutrino Observatory is an array of 5,160 digital optical modules (DOMs) instrumenting 1 km$^3$ of clear Antarctic ice at the geographic South Pole \cite{Aartsen:2016nxy}. The DOMs are deployed on 86 strings, with 60 DOMs per string, at depths ranging from \SI{1,450}{\meter} to \SI{2,450}{\meter} below the surface of the ice sheet. The strings have an average nearest-neighbor separation of \SI{125}{\meter}, and on each string the DOMs are spaced vertically by \SI{17}{\meter}. The detector records the Cherenkov light produced by cosmic ray muons and the charged particles created by neutrinos interacting in the ice, and uses the relative timing of direct and indirect Cherenkov photons to reconstruct the energies and arrival directions of the cosmic muons and neutrinos.

The sparse geometry of IceCube is optimized to reconstruct the arrival directions of neutrinos with energies between \SI{10}{\giga\electronvolt} and \SI{10}{\peta\electronvolt} \cite{Aartsen:2016nxy}, so the detector cannot be used to reconstruct tracks from the $\mathcal{O}(10~\text{MeV})$ neutrinos produced in CCSNe. However, IceCube is sensitive to the positrons generated during the inverse beta decay of $\bar{\nu}_e$, which arrive over $\sim10$\,s during the accretion and cooling phases of a CCSN \cite{Abbasi:2011ss}. The optical absorption length in the ice is $>100$\,m, so the effective volume of each DOM is several hundred cubic meters. The positrons produced in the ice during the $\bar{\nu}_e$ burst are therefore observable as a rise in the background count rates in the DOMs over a period of \SI{0.5}{\second} to \SI{10}{\second}.

Depending on the mass of the stellar progenitor, IceCube will record $10^5$ to $10^6$ photoelectrons from a supernova located \SI{10}{\kilo\parsec} from Earth, enabling detailed measurements of the neutrino light curve \cite{Abbasi:2011ss, Cross:2019jpb}. If the light curve has significant time structure, such as a cutoff in the $\bar{\nu}_e$ flux due to the creation of a black hole, IceCube can provide limited pointing toward the source with a resolution $\Delta\psi\approx10^\circ-100^\circ$, depending on distance and model assumptions. The DOM counts can also constrain the shape and average energy the $\bar{\nu}_e$ spectral energy distribution with a resolution of $\Delta E/E\>25\%$ \cite{Koepke:2017req}.

Construction of IceCube finished in 2011, and since 2015 the trigger-capable live time of the detector has averaged 99.7\% per year. IceCube thus provides a critical high-uptime monitor for CCSNe in the Milky Way and Magellanic Clouds. A realtime monitor tracks the photoelectron rates in the IceCube DOMs in with a resolution of \SI{2}{\milli\second}. The monitor performs a moving-average search for a collective rise in the DOM rates in sliding bins of \SI{0.5}{\second}, \SI{1.5}{\second}, \SI{4}{\second}, and \SI{10}{\second} \cite{Abbasi:2011ss}, as well as an unbinned search for significant changes in the collective count rate \cite{Cross:2017icrc}. When a statistically significant increase in the DOM counts is observed, an alert is sent to SNEWS with a latency of $\sim$ \SI{5}{\minute}. The software also buffers the full DOM waveforms in a \SI{600}{\second} window around the time of the alert. The buffered waveforms are not limited by the 2 ms resolution of the online system, and can be used for detailed model fits of the CCSN light curve \cite{Koepke:2017req}.

The IceCube detector is currently being upgraded with two new strings instrumented with multi-anode DOMs \cite{Ishihara:2019aao, Classen:2019icrc}, which if deployed in large numbers can improve constraints on the CCSNe neutrino average energy to $\Delta E/E\approx5\%$ \cite{Lozano:2018neutrino}. A build-out of the detector to $\SI{8}{km^3}$ and $\sim10,000$ optical modules (IceCube Gen-2), planned for the late 2020s, will more than double the photocathode area in the detector \cite{Aartsen:2019swn}.

\subsubsection{KM3NeT}

is a research infrastructure under construction in the Mediterranean Sea. It comprises two deep-sea neutrino detectors, ORCA, located offshore Southern France, optimised for neutrino oscillation studies and ARCA, located offshore Sicily, optimised for neutrino astronomy. They respectively focused on neutrino oscillations and astrophysics. KM3NeT detectors are arrays of multi-PMT digital optical modules (DOMs). The two sites will be instrumented with over 6,000~DOMs for a total of about 200,000~PMTs. KM3NeT DOMs are arranged in vertical detection lines of 18~DOMs, which are deployed at depths between 2,500-\SI{3,400}{\meter} to form a three-dimensional array~\cite{LoIKM3NeT}.

KM3NeT sensitivity to CCSN neutrinos on the tens-of-MeV energy scale is mainly achieved through the detection of inverse beta decay interaction products in the proximity of the DOMs. A supernova neutrino burst is expected to produce an increase in the coincidence rate between different PMTs on the same optical module. The number PMTs detecting a photon in coincidence can be used to discriminate the signal from the natural backgrounds, namely bioluminescence, $^{40}$K-dominated radioactive decays and atmospheric muons~\cite{MUONDEPTH}.

To achieve the best detection sensitivity, a selection based on coincidences with at least six hit PMTs is defined. The background from atmospheric muons is reduced by rejecting events which are correlated over multiple DOMs. This reduces the expected background rate below \SI{1}{\hertz} per detection unit. With this selection, the signal expectation for the accretion phase of a CCSN exploding at \SI{10}{\kilo\parsec} is between 50 and 250 events for a 115-lines instrumented block. This selection is used for the sensitivity estimation and implemented in the real-time KM3NeT supernova trigger.

The KM3NeT detectors will be sensitive beyond the Galactic Centre, reaching the coverage of $\sim$90-95\% of progenitors in the Galaxy. Both the online and offline searches are applied within a time window of \SI{500}{\milli\second}, covering the accretion phase. For the real-time trigger, data is sampled with a frequency of \SI{10}{\hertz}. The expected detection horizon for a false alert rate compatible with the SNEWS requirement varies between 13 and \SI{26}{\kilo\parsec}, depending on the considered flux model.

For the purpose of CCSN astrophysics studies, the signal statistics can be improved using a selection of all coincidences (two or more hit PMTs on a DOM) at the cost of a lower purity. With a complete detector at each site (one instrumented block in ORCA and two in ARCA), KM3NeT is expected to collect 20-50,000 events (see Table~\ref{t:detev}). It will be able to study the neutrino light-curve for signatures of time-dependent patterns, the determination of the arrival time of the burst and the characterisation of the neutrino spectrum~\cite{ICRCprocSNKM3}.

The effective mass of a KM3NeT 115-lines instrumented block is 40-\SI{70}{\kilo\tonne} for all coincidences and 1-\SI{3}{\kilo\tonne} for coincidences with six or more hit PMTs.

The KM3NeT real-time supernova search is currently operational with a dedicated data processing pipeline, running on the already connected detection units. Alerts can be sent within \SI{20}{\second} from the generation of the corresponding data off-shore. The triggering capability under SNEWS requirements already reaches sensitivity to progenitors closer than 4.5-\SI{8.5}{\kilo\parsec}. This horizon will increase in the near future with the deployment of new detection lines.

\subsection{Scintillator Detectors}

\subsubsection{Baksan}
\label{sec:baksan}
The Baksan Underground Scintillation Telescope was one of the detectors which observed neutrinos from SN1987A~\cite{Alekseev:1987ej} and is still operational, having continuously watched for SN neutrinos from 1980 till the present~\cite{Novoseltsev:2019gdt}.  The experiment is at a depth of \SI{850}{\mwe} in the Baksan Valley, Kabardino-Balkari, Russia.  The experiment's \SI{330}{\tonne} of liquid scintillator is primarily sensitive to positrons produced by IBD, and is divided into two independent sub-detectors operating in coincidence with each other.  The detector elements are grouped by background rate (and thus threshold), with ``D1'' having \SI{130}{\tonne} at a \SI{8}{\mega\electronvolt} threshold and a background rate of \SI{0.02}{\hertz}, and ``D2'' \SI{110}{\tonne} at \SI{10}{\mega\electronvolt} and \SI{0.12}{\hertz}.  This arrangement further reduces the already low background rate, contributing to the detector's long-term stability and $>$90\% uptime.  In event of a supernova, the remaining \SI{90}{\tonne} of scintillator would also record events, but this section has a background rate of \SI{1.4}{\hertz} so is excluded from the online SN trigger.  The detector and its supernova trigger are detailed in  ~\cite{Novoseltsev:2019gdt}.  A future \SI{10}{\kilo\tonne} detector is under development, to be placed at a depth of \SI{4800}{\mwe}~\cite{Petkov:2020ubs}.

\subsubsection{LVD}
The Large Volume Detector (LVD) is a modular \SI{1}{\kilo\tonne} liquid scintillation detector and consists of an array of 840 counters located underground  (minimal depth \SI{3000}{\mwe}), in the INFN Gran Sasso National Laboratory (Italy). It has been  designed to study  neutrinos from CCSN mainly through IBD interactions \cite{Aglietta:1992dy}. The detector modularity allows the experiment to achieve a very high duty cycle, essential in the search of unpredictable sporadic events, typically greater than 99\% over more than 20 years. 
LVD has been in operation since 1992, after a short commissioning phase, with an active mass increasing from \SI{300}{\tonne} to \SI{1000}{\tonne} at the time of completion in 2001. The experiment is approaching the decommissioning phase planned for the end of 2020, after almost 30 years of continuous operations.

During these years, as shown in~\cite{Agafonova:2015}, LVD has been monitoring  the entire Galaxy. The number of expected neutrino events in LVD from a CCSN have been evaluated via a parameterised model based on a maximum likelihood procedure based on the SN1987A data assuming energy equipartition and normal mass hierarchy for neutrino oscillations~\cite{Pagliaroli:2009}. The resulting average $\bar\nu_e$ energy is 
$\bar E_{\bar\nu_e}=$ \SI{14}{\mega\electronvolt}, for a $E_b = \SI{2.4e53}{\erg}$  total energy SN. At a reference distance of \SI{10}{\kilo\parsec}, we expect a total of 300 events, 88\% due to IBD.
The burst search, i.e. the search of  cluster of events in the time series of triggers, is performed on a pure statistical basis  as explained in detail in  \cite{Agafonova:2015}.

LVD is a founding member of the SNEWS project and it has been participating in the on-line operations since the beginning  in 2005 together with the SuperK and SNO detectors~\cite{Antonioli:2004zb}. 

\subsubsection{Borexino} is an ultra-pure liquid organic scintillation detector designed to study solar neutrinos \cite{Alimonti:2008gc} in real time. There are two main interaction channels, namely elastic scattering on electrons and inverse beta decay, by which neutrinos are seen. In case of a supernova, observation of other processes becomes important including elastic scattering on protons and various reactions on carbon. Nevertheless it is reasonable due to low statistics to organize alarm datagram generation for SNEWS based on searching for IBD event bursts and/or point-like event excesses. 
	
Despite the relatively small mass of the target (\SI{278}{\tonne}) and therefore the quite limited event numbers from nearby supernovae compared to other modern detectors, Borexino is a rather sensitive apparatus because of very low background levels. If the Borexino SN alarm is the result of searching for an excess of point-like events, the respective mean background rate is $\mathcal{O}$(\SI{0.01}{\hertz}) above \SI{1}{\MeV}. Since the alarm is generated based on the IBD analysis, the background level is even lower. For example, only 154 golden IBD candidates were found in a live time of approximately 9 years in the last geo-neutrino investigation with Borexino \cite{Agostini:2019dbs}. Such low background conditions were achieved thanks to a number of features of the detector, its special location and applied methods of data analysis. To begin with, Borexino is placed in the underground laboratory (LNGS) under a thick layer of rocks that provide the shielding against cosmogenic background equivalent \SI{3800}{\mwe} and therefore the muon flux is decreased up to \SI{3.432 \pm 0.003 e-4}{\per\square\meter\per\second}~\cite{Agostini:2018fnx}. 
There are no nuclear power plants in Italy and the mean weighted distance of the reactors from the LNGS site is more than \SI{1000}{\kilo\meter}, resulting in a low antineutrino background in the detector. Other reasons for the low background rates are concentric multilayer shielding and ultra-radiopurity of all materials. The complex selection and refinement of the detector materials, accurate assembling guarantee extremely low levels of $^{238}$U and $^{232}$Th 
contamination that are less than $<$ \SI[per-mode=symbol]{9.4e-20}{\gram\per\gram} ($95\%$ C.L.) and $<$ \SI[per-mode=symbol]{5.7e-19}{\gram\per\gram} ($95\%$ C.L.) respectively \cite{Agostini:2019dbs}. 
	
The light yield in Borexino is approximately \SI{500}{photoelectrons} (p.e.) per \SI{1}{\mega\electronvolt} and the energy resolution scales as $\sim 5\%/\sqrt{E~(\text{MeV})}$ \cite{Alimonti:2008gc}. Borexino is a position sensitive detector. The absolute errors of the reconstructed coordinates $x$, $y$, $z$ for a point-like event in the target are virtually the same and depend on the energy as $\sim$\SI{10}{\centi\meter}/$\sqrt{E~(\text{MeV})}$. To classify the events,  pulse shape discrimination methods (PSD) are actively applied. 
	
Since the main DAQ of the Borexino detector (``LABEN'') was designed for spectroscopy of solar neutrinos with energies in the sub-MeV energy range, a special additional data acquisition system based on flash ADCs (``FADC'') was designed to record neutrinos with energies up to $\sim$\SI{50}{\mega\electronvolt} from supernovae. Both systems are working and will continue to function until the end of the experiment.
	
Borexino joined the SNEWS community in 2009. Originally there were two independent modules for the LABEN data analysis and generation the alarms: the Echidna Online supernova monitor and the Princeton supernova monitor. Currently only the latter is operational. The Borexino experiment is approaching its end. The data taking is planned to stop in 2021.

\subsubsection{KamLAND}

Kamioka Liquid Scintillator Anti-Neutrino Detector~(KamLAND) is an experiment to detect $\bar{\nu}_e$ signals using \SI{1000}{\tonne} of ultra pure liquid scintillator. The KamLAND detector occupies the former site of the Kamiokande experiment in the Kamioka mine (\SI{2700}{\mwe} overburden). A spherical balloon of \SI{13.0}{\meter} diameter is suspended in the center of the KamLAND detector and filled with \SI{1000}{\tonne} (\SI{1171\pm25}{\cubic\meter}) of ultra pure LS. Scintillation photons are detected by 1,325 fast 17-inch aperture Hamamatsu PMT custom-designed for KamLAND, and 554 20-inch PMTs inherited from Kamiokande. The PMT array are attached inside on the spherical stainless steel tank of 18.0m diameter. The outside of the tank is a water Cherenkov detector for veto with approximately \SI{3000}{\cubic\meter} of pure water and 140 high-QE PMTs~\cite{Ozaki:2016fmr}. Event vertex and energy reconstruction is based on the timing and charge distributions of scintillation photons. In 2013, they are estimated to be $\sim 12~\text{cm}/\sqrt{E~(\text{MeV})}$ and $6.4~\%/\sqrt{E~(\text{MeV})}$, respectively~\cite{Gando:2013nba}. 

The main channel for supernova neutrinos in KamLAND is $\bar{\nu}_e$ via inverse beta decay. 200--300 events are expected from a galactic supernova at \SI{10}{\kilo\parsec}. Because of delayed coincidence measurements with temporal and spatial correlation between the positron and neutron events resulting from IBD, very low backgrounds for $\bar{\nu}_e$ are possible. KamLAND uses this channel for supernova alarms. If the online process finds two IBD events during \SI{10}{\second}, KamLAND will send an supernova alert to the SNEWS server. 

The proton recoil~($\nu + p \to \nu + p$) is one of the unique features in LS detectors~\cite{Beacom:2002hs}. This channel provides a chance to study the total energy, temperature~\cite{Beacom:2002hs} and reconstruction of the $\nu_x$ spectrum~\cite{Dasgupta:2011wg}. Since this channel does not have background rejections like the delayed coincidence measurements, low-background measurements are required. KamLAND had performed a purification campaign from 2007 to 2009, enabling measurements of supernova neutrinos using proton recoils. In the current configuration of the KamLAND trigger, the expected number of events is about 60 with assumptions of total luminosity of \SI{3e53}{\erg} and $\langle E_{\nu_e} \rangle = \SI{12}{\mega\electronvolt}$, $\langle E_{\bar{\nu}_e} \rangle = \SI{15}{\mega\electronvolt}$, and $\langle E_{\nu_x} \rangle = \SI{18}{\mega\electronvolt}$. 

It is also possible to detect neutrinos emitted before collapse (``pre-supernova neutrinos'') using IBD. KamLAND is able to detect pre-supernova neutrinos from nearby stars, such as Betelgeuse and Antares, before their cores collapse. The background rate of low energy IBD in KamLAND is typically 0.1 event/day. This strongly depends on the reactor on/off status in Japan since the main background is neutrinos emitted from reactors. Even if the expected cumulative number of IBD events from a pre-supernova in a \SI{48}{\hour} period is only a few, the background is low enough that this is a statistically significant detection~\cite{Asakura:2015bga}. With a latency of about 25~minutes, this detection significance is open for authorized users. If a detection significance is larger than $\SI{3}{\sigma}$ (approximately three IBD events in the last \SI{48}{\hour}), the KamLAND collaboration has a special meeting to discuss the result. Based on that discussion, KamLAND will send a message to The Astronomer's Telegram. The connection from KamLAND to The Gamma-ray Coordinates Network~(GCN) and SNEWS~2.0 will be implemented. 

KamLAND is planning new electronics and DAQ systems which will improve the latency of supernova alarms. To better measure proton recoils with a low energy threshold, the trigger threshold will be changed just after the  detection of the first 10 events.

Unfortunately, supernova neutrino observations in KamLAND has a small conflict to KamLAND-Zen, an experiment to search for neutrino-less double-beta decay of $^{136}$Xe with the existing KamLAND detector. In 2011, a inner balloon with a \SI{3.08}{\meter} diameter was installed in the center of the KamLAND detector~\cite{KamLANDZen:2012aa}. Xe-loaded LS was filled in the inner balloon. Because of double-beta decays of $^{136}$Xe and backgrounds from the balloon materials, an additional cut for  $\bar{\nu}_e$ analysis~\cite{Gando:2013nba} is applied. In 2018, an updated inner balloon with a \SI{3.84}{\meter} diameter was installed. Improvements of the analysis methods are being studied to maximize the sensitivity to supernova neutrinos. 

\subsubsection{JUNO}

The Jiangmen Underground Neutrino Observatory (JUNO) is a multi-purpose neutrino experiment~\cite{An:2015jdp}. It primarily aims to determine the neutrino mass ordering using reactor antineutrinos. The facility is under construction in Jiangmen city, Guangdong province, China and expected to start data taking in 2022. The detector consists of a central detector, a water Cherenkov detector and a muon tracker, deployed in a laboratory \SI{700}{\meter} underground  ($\sim \SI{1800}{\mwe}$). The central detector is a \SI{20}{\kilo\tonne} liquid organic scintillation (LS) detector with a designed energy resolution of $3\%/\sqrt{E({\rm MeV})}$, with a goal to reach a rather low detection energy threshold of \SI{0.2}{\MeV} with a conservative PMT dark noise rate of \SI{50}{\kilo\hertz}~\cite{Fang:2019lej}, limited by the intrinsic radioactive background mainly from $^{14}{\rm C}$ in the LS. 

Supernova detection is one of the major purposes for JUNO. Given the large fiducial mass and low energy threshold of the LS detector, JUNO is able to register all flavors of $\mathcal{O}$(\SI{10}{MeV}) supernova burst neutrinos with relatively high statistics. For a typical CCSN at \SI{10}{\kilo\parsec}, there could be $\sim$\SI{5000}{{\rm events}} from the inverse beta decay (golden) channel, $\sim$\SI{2000}{{\rm events}} from neutrino-proton elastic scattering, more than \SI{300}{{\rm events}} from neutrino-electron elastic scattering, as well as charged current and neutral current interactions of neutrinos on  $^{12}{\rm C}$ nuclei expected in JUNO~\cite{Lu:2016ipr}. IBD events with anisotropic prompt and delayed signals could determine the supernova pointing with an uncertainty of $\mathcal{O}$(\ang{10})~\cite{Apollonio:1999jg}. By combining all these channels, JUNO has the unique opportunity to provide  flux information and to reconstruct the energy spectrum  of $\nu_{x}$ ($\nu_{\mu}$, $\nu_{\tau}$ and $\overline{\nu}_{\mu}$, $\overline{\nu}_{\tau}$) supernova burst neutrinos~\cite{Li:2017dbg, Li:2019qxi}.  

For the nearby galactic CCSN within $\sim$\SI{1}{\kilo\parsec}, JUNO is also sensitive to the $\mathcal{O}$(\SI{1}{MeV}) pre-supernova neutrinos via inverse beta decay and neutrino-electron elastic scattering, thanks to its low energy threshold. The IBD events could be clearly distinguished using the coincidence signals, while the neutrino-electron interaction of pre-supernova suffers from the high rate of background induced by the radioactivity of detector materials and cosmogenic isotopes. Therefore, the pre-supernova could be monitored online with the IBD candidates in JUNO, including possibly pointing information once a large number of the IBD events are accumulated~\cite{Li:2020gaz}. In addition, \SI{2}{{\rm GB}} DDR3 memory will be embedded on the global control units in the front-end electronics design of JUNO~\cite{Pedretti:2018xjc}, to buffer the rapid increase of events from supernova burst neutrinos from a nearby galactic CCSN. 

Currently the real-time monitor strategy for both supernova burst neutrinos and pre-supernova neutrinos is under progress in JUNO, which consists of prompt monitor on hardware and DAQ monitor. The experiment is capable of providing the supernova alerts as well as the pre-supernova ones to SNEWS in the near future. 

\subsubsection{SNO+}

The SNO+ experiment is a liquid scintillator detector situated \SI{2092}{\meter} underground at SNOLAB~\cite{Andringa16}.  It reuses much of the infrastructure of the SNO detector, which first demonstrated flavor conversion in $^8$B solar neutrinos, including the \SI{12}{\meter}-diameter acrylic vessel (AV) containing the active material, and some 9,300~PMTs arranged in a geodesic support structure with a diameter of approximately \SI{17.8}{\meter}.  The flat rock overburden provides \SI{5890\pm 94}{\mwe} shielding and reduces the cosmic muon flux to 63~muons per day through an \SI{8.3}{\meter}-radius circular area.

The main goal of SNO+ is to observe neutrinoless double-$\beta$ decay of the $^{130}$Te nucleus.  Tellurium will be dissolved in a cocktail of linear alkylbenzene (LAB) and 2,5-diphenyloxazole (PPO).  The expected light yield is over 400 PMT hits per MeV.  In addition to the new infrastructure needed to handle and purify liquid scintillator, upgrades have been made to the buffered readout electronics, cover gas system, and calibration systems using LED, laser, and radioactive sources. The trigger readout window is roughly 400ns and is essentially deadtime-less. SNO+ began taking data with an AV filled with water in May 2017. Over the course of 2020, SNO+ is replacing the water in the AV with \SI{780}{\tonne} of scintillator, and soon after expects to load the detector with 0.5\% (by mass) of tellurium, resulting in 1330kg of the target isotope.  Techniques for increasing the Te load are promising.

As with other liquid scintillator experiments, SNO+ will be sensitive to IBDs and neutrino elastic scatters off electrons. The high IBD efficiency, combined with a very low background rate, will be exploited to add to the SNEWS pre-supernova alert. In addition, the low background rate makes it possible to observe elastic scattering off protons. Though the energy deposit from such scatters is highly quenched, they may represent nearly half of the neutrinos observed from a supernova at SNO+, and are an important channel for flavor-agnostic neutral current interactions. Table~\ref{tab:snoplusyield} illustrates yields from different interaction channels for a lower, conservative light yield of 200 hits per MeV ({\it i.e.}, less than half of the expected light yield given above). Efforts are also underway to investigate whether different fluors can help distinguish directional Cherenkov light ({\it e.g.}, \cite{Biller:2020uoi}).

\begin{table}[htbp]
\begin{center}
\begin{tabular}{lc}
\hline
Reaction & Yield \\ \hline
NC: $\nu p\rightarrow \nu p$ & $429.1\pm 12.0$ \\ \hline
CC: $\overline{\nu}_e p \rightarrow e^+n$ & $194.7\pm 1.0$ \\ \hline
CC: $\overline{\nu}_e\;^{12}{\rm C}\rightarrow e^+\;^{12}{\rm B}_{gs}$ & $7.0\pm 0.7$ \\
CC: $\nu_e\;^{12}{\rm C}\rightarrow e^-\;^{12}{\rm N}_{gs}$ & $2.7\pm 0.3$ \\
NC: $\nu\;^{12}{\rm C}\rightarrow \nu'\;^{12}{\rm C}^\ast(15.1\;{\rm MeV})$ & $43.8\pm 8.7$ \\ \hline
CC/NC: $\nu \;^{12}{\rm C} \rightarrow \;^{11}{\rm C}$ or $\;^{11}$B$+X$ & $2.4\pm 0.5$ \\ \hline
$\nu e$ elastic scattering & 13.1 \\ \hline
\end{tabular}
\caption{Illustrative neutrino interaction yields, from \cite{Andringa16}, for \SI{780}{\tonne} of LAB-PPO and a supernova at \SI{10}{\kilo\parsec}. The supernova model releases $3\times 10^{53}$ erg equally partitioned among all neutrino flavors and considers no flavor changing mechanisms. The NC $\nu p$ yield will be reduced to $118.9\pm 3.4$ if the trigger threshold is set at \SI{0.2}{\mega\electronvolt}. The uncertainties quoted here include only cross section uncertainties (the Standard Model cross section uncertainty on $\nu e$ elastic scattering is less than 1\%).}
\label{tab:snoplusyield}
\end{center}
\end{table}

\subsubsection{NOvA}
The NOvA experiment consists of two segmented liquid scintillator detectors. They are functionally identical, and mainly differ by mass and location. The \SI{300}{\tonne} Near Detector is located at Fermilab, near Chicago, at a depth of \SI{100}{\meter}. The \SI{14}{\kilo\tonne} Far Detector is located near Ash River, Minnesota, and sits on the Earth's surface with a modest barite overburden. The detectors are composed of hollow extruded PVC cells \SI{3.9}{\centi\meter} $\times$ \SI{6.6}{\centi\meter} in cross-sectional area, which are connected together to form planes. The planes are glued together to form the full body of the detector, alternating between vertical and horizontal orientations to allow for reconstruction in three dimensions. The PVC cells are filled with a mixture of mineral oil and pseudocumene. Scintillation photons produced by the passage of charged particles through the cells bounce off of the highly-reflective cell walls until they are absorbed by a wavelength-shifting fiber. The fiber is looped through the length of the cell, with both ends terminating on the same channel of an avalanche photodiode.

Inverse beta decay ($\bar{\nu}_e + p \rightarrow e^+ + n$) is the most common interaction mode for supernova neutrinos in the NOvA detectors, followed by elastic scattering on electrons, and neutral current interactions on carbon nuclei. Detection threshold is $\sim$\SIrange{8}{15}{\mega\eV}, depending on how close to the readout the energy deposition occurred, meaning many of the expected interactions will be detectable.

There are several challenges to reconstructing supernova neutrinos with NOvA. Due to its size, the Far Detector will see many more neutrino interactions than the Near Detector, but it is on the surface and subject to a cosmic ray muon rate of $\sim$\SI{148}{\kilo\hertz}. It is easy to identify and veto the muons themselves with software, along with the Michel electrons that are often produced at the end of stopping tracks, but identifying the spallation products they leave in their wake is more difficult.

The NOvA DAQ employs a buffered readout system that allows for continuous readout of the detector. In its current configuration, the DAQ writes \SI{45}{\second} of continuous data to disk in the event of a supernova trigger. NOvA subscribes to SNEWS and issues a supernova trigger for any SNEWS alert. A data-driven supernova trigger has also been developed and running on both detectors since November 2017~\cite{novaSN}. This data-driven trigger removes known backgrounds and groups hits into candidate IBD interactions. A supernova will manifest as a sudden increase in the rate of IBD candidates, which then slowly returns back to baseline.

A time series of IBD candidates is constructed in real time and compared against two hypotheses, one which represents the background only and another which represents a signal superposed on the background. A log likelihood ratio is computed, which is then converted into a signal significance. Several models for the signal shape, including a flat one, have been tested for this method, and they all show similar discrimination power. With this method, NOvA can identify a potential supernova signal in a model-independent way. A trigger is issued when the signal significance exceeds a threshold, which is currently chosen to be 5.6~$\sigma$ and corresponds to a SNEWS-inspired false alarm rate of one per week.

\subsection{Lead-Based Detectors}

\subsubsection{HALO}
The Helium And Lead Observatory (HALO) utilizes lead-based detection, and has been operating at SNOLAB since 2012. HALO consists of \SI{79}{\tonne} of lead, instrumented with $^3$He neutron counters. These counters have a very low background rate, which allows sufficient trigger discrimination for HALO to be used in the SNEWS network. 

Lead-based neutrino detection is possible through charged current or neutral current interactions with lead nuclei. The charged current interactions are: $\nu_e + \mbox{\textsuperscript{208}Pb} \rightarrow \mbox{\textsuperscript{207}Bi} +n+e^-$ and  $\nu_e + \mbox{\textsuperscript{208}Pb} \rightarrow \mbox{\textsuperscript{206}Bi} +2n+e^-$. Their kinematic threshold energies (weighted over the isotopic abundances) are \SI{10.2}{\mega\electronvolt} and \SI{18.1}{\mega\electronvolt}, respectively. The neutral current interactions: $\nu_e + \mbox{\textsuperscript{208}Pb} \rightarrow \mbox{\textsuperscript{207}Pb} +n$ and  $\nu_e + \mbox{\textsuperscript{208}Pb} \rightarrow \mbox{\textsuperscript{206}Pb} +2n$ have weighted threshold energies  \SI{7.4}{\mega\electronvolt} and \SI{14.4}{\mega\electronvolt}, respectively. Because these interactions have different cross sections and thresholds the number of $1n$ and $2n$ events depends of the incident neutrino energy. The neutrons that result from these interactions can travel some distance through the lead; they are thermalized by collisions with the lead and with the polyethylene moderator prior to being captured by $^3$He gas in the proportional counters. Lead-based neutrino detectors are sensitive primarily to electron neutrinos, because $\bar{\nu}_e$ charged-current reactions are strongly suppressed via Pauli blocking. 

The relatively small lead mass of HALO limits its effectiveness for studying supernovae from distances beyond $\sim$ \SI{5}{\kilo\parsec}. 

\subsubsection{Future Lead-Based Detectors}

HALO-1kT is a proposed upgrade to HALO to be sited at LNGS which would leverage \SI{1000}{\tonne} of lead from the decommissioned OPERA experiment in order to create a low cost and low maintenance neutrino detector. The much larger target mass will greatly increase the sensitivity to supernova neutrinos. The lead will be instrumented with a $28\times28$ array of $^3$He detectors, comprising \SI{10000}{\liter\atm} of $^3$He. The greatly increased target mass should allow detection of supernova at much greater distances. The major advantage of these lead-based detectors is the very low maintenance and high livetime, which could allow them to maintain continuous operation for decades. Supernova neutrino detection would be a primary mission for HALO-1kT, with the expectation of a multiple decade long uninterrupted search for supernovae.

RES-NOVA~\cite{Pattavina:2020cqc} aims to obtain a large SN signal in a lead detector using a small amount of very low-background archaeological lead, allowing the lowering of the threshold and giving sensitivity to the higher-rate \cevns\ interactions.

\subsection{Liquid Noble Dark Matter Detectors}

Noble liquids are one of the leading targets for dark matter direct detection efforts, assuming that dark matter is composed of WIMPs (Weakly Interacting Massive Particles). In the currently planned future noble liquid detectors, the active volume of liquid is hosted in a dual-phase time projection chamber (TPC) containing also a gas pocket of the same target material above the active volume.  The two volumes are separated by a drift-electron extraction grid positioned just below the liquid surface, thus establishing the uniform drift region in the active liquid volume and the electroluminescene in the gaseous volume.

When the incident particle, WIMP as well as supernova neutrinos, scatters on the noble liquid atom, the recoil energy goes into phonons (not detected), scintillation light that is detected by the photosensors and called the S1 signal, and ionization electrons. The ionization electrons are drifted by an electric field parallel to the axis of the chamber towards the grid at the top of the detector, where they are extracted into a region of the gaseous noble element and accelerated, emitting a delayed photon signal from electroluminescence, called S2 and also recorded by the photosensors.  As the average neutrino energy is $\sim$\SI{10}{\mega\electronvolt}, the dominant cross section in liquid nobles is from the \cevns\ interaction, which allows for a flavor-insensitive detection. Low thresholds achievable with S2-only signals allow for detection of Type~Ia supernova neutrinos~\cite{Raj:2019sci} in addition to the core-collapse neutrinos which are the usual supernova neutrino burst targets (Sec.~\ref{sec:typeIa}).

\subsubsection{Global Argon Dark Matter Collaboration}
The Global Argon Dark Matter Collaboration's (GADMC) aim is the direct detection of WIMPs in liquid argon (LAr). Strong exclusion limits for WIMP-nucleon spin-independent interactions have already been set by the running detectors, DarkSide-50 (\SI{50}{\kilo\gram} of LAr)~\cite{Agnes:2018vl} and DEAP-3600 (\SI{3279}{\kilo\gram} of LAr)~\cite{Ajaj:2019jk}. In order to improve the sensitivity, a future detector's target mass must be scaled up to tonnes and be extremely radiopure. Hence, the upcoming multi-tonne detectors, DarkSide-20k (DS-20k) (\SI{47}{\tonne} LAr) and Argo (\SI{300}{\tonne} LAr), will be built with radiopure materials including underground argon as the detection target~\cite{Agnes:2016fz}, and use silicon photomultipliers (SiPM) rather than PMTs for photon detection. Based on an analysis done with an input supernova neutrino flux from the Garching group simulation for a \SI{27}{\solarmass} progenitor mass following the LS220K equation of state~\cite{Mirizzi:2015eza}, DS-20k will be capable of observing $\sim$340 neutrino events after selection cuts at a distance of \SI{10}{\kilo\parsec}.  This shows that DS-20k will be capable of observing the neutrino signal from such a core-collapse supernovae with more than $\SI{5}{\sigma}$ significance up to \SI{40}{\kilo\parsec}\cite{Agnes:2020pbw}.

\subsubsection{Xenon}

The LZ and XENON Collaborations aim for the direct detection of WIMPs, but using liquid xenon (LXe) TPCs \cite{Akerib:2019fml,Aprile:2017aty}. The XENON1T experiment, with about \SI{2}{\tonne} of active target mass, has set the most stringent limits on WIMP cross-sections for masses above \SI{6}{\giga\electronvolt} \cite{Aprile:2018dbl}. It has also measured the half life of \(^{124}{\rm Xe}\) decaying via two-neutrino double electron capture, the rarest decay process to have ever been directly measured \cite{XENON:2019dti}.Through its lifetime, XENON1T has been directly connected with SNEWS in order to receive alarms and promptly save any related data of interest for later analysis \cite{ Aprile:2019cee}.

The future LZ and XENONnT detectors will have about 6 to \SI{7}{\tonne} of active target mass \cite{Akerib:2019fml,Aprile:2020vtw}. Detectors of this size are sensitive to extremely rare decay processes such as neutrinoless double beta decay \cite{Akerib:2019dgs} and will be able to observe the neutrino signal from a core-collapse supernova with a \SI{27}{\solarmass} progenitor happening anywhere in the Milky Way with $\SI{5}{\sigma}$ significance \cite{Lang:2016zhv}.  Such a supernova at \SI{10}{\kilo\parsec} will produce 350 neutrino interactions in total in the active volume of LZ~\cite{Khaitan:2018wnf}.  More specifically, for a threshold of $\sim$3 detected electrons, the number of supernova neutrino events for the \SI{27}{\solarmass} supernova progenitor with the LS220 EoS is \SI{17.6}{events/tonne} over the first seven seconds post bounce \cite{Lang:2016zhv}. Both the LZ and XENON collaborations are currently developing a real-time supernova trigger to be incorporated in their detectors and data framework, aiming to actively contribute to SNEWS in the near future~\cite{Khaitan:2018wnf}.

A next-generation xenon detector, such as the one proposed by the DARWIN collaboration~\cite{Aalbers:2016jon}, will further increase the sensitivity to core-collapse supernovae. It is expected to discern with $\SI{5}{\sigma}$ significance such an event from a \SI{27}{\solarmass} progenitor beyond \SI{60}{\kilo\parsec} and possibly distinguish between different supernova models and progenitor masses for a close supernova event~\cite{ Lang:2016zhv,Aalbers:2016jon}.

\subsection{Liquid Argon Time Projection Chamber Neutrino Detectors} 

The detection of a supernova neutrino burst through the $\nu_{e}+{\rm ^{40}Ar}\rightarrow e^{-}+{\rm ^{40}K^{*}}$ channel is especially interesting due to the enhanced sensitivity to the electron-neutrino flux, which is complementary to the dominant electron-antineutrino flux from other detectors. Liquid argon time projection chambers (LArTPCs) drift ionization charge created by particle tracks through liquid argon volumes; the drifted charged is measured in a 2D plane projection using various possible readout methods.  The relative drift arrival times of the measured charges then provide the third dimension for track reconstruction. Liquid argon also scintillates at \SI{128}{\nano\meter}, and these fast scintillation photons can also be measured to determine position along the drift direction as well as a measure of energy deposition.  LArTPCs do not require energies above the Cherenkov threshold and with sufficiently fine-grained readout can provide precision particle track reconstruction.

The future DUNE very large LArTPC experiment has the detection of neutrinos from a core-collapse supernova within our galaxy as one of its primary science goals, but the data taking with the first module is not expected before 2026. In the meantime, a series of smaller LArTPCs already or soon to be in operation represent our best chance to detect supernova neutrinos in argon. Among these, the Fermilab Short-Baseline Neutrino Program experiments MicroBooNE \cite{Acciarri:2016smi} (taking neutrino beam data since October 2015) and SBND \cite{Antonello:2015lea} (expected to begin data taking in 2021) are pioneering a novel approach for the detection of the supernova neutrinos. Due to their active masses (\SI{90}{\tonne} for MicroBooNE, \SI{112}{\tonne} for SBND), only tens of events are expected for a canonical core-collapse supernova at \SI{10}{\kilo\parsec}. Their location near surface exposes them to a large flux of cosmic rays of a few kHz, making triggering on the supernova neutrinos very challenging. For these reasons, the MicroBooNE and SBND detectors, rather than sending alerts to SNEWS, are instead SNEWS consumers, using the ``Gold'' alert as DAQ trigger. In order to accommodate the latency of the SNEWS alert, both experiments feature a continuous readout of the TPC~\cite{Abratenko:2020hfy}, which is written to disk for several hours and subsequently deleted if no SNEWS alert is received.

\subsubsection{DUNE}

The Deep Underground Neutrino Experiment (DUNE) is a \SI{40}{\kilo\tonne} (fiducial mass) LArTPC detector, made of four \SI{17}{\kilo\tonne} (\SI{10}{\kilo\tonne} fiducial) modules to be constructed underground in South Dakota. There are two designs under consideration for the DUNE far detector TPCs:  a ``single-phase'' design that features horizontal drift along with readout comprising one charge-collection wire plane and two induction wire plane with \SI{5}{mm} pitch, and a ``dual-phase'' design that features vertical charge drift and charge amplification and collection at a top gas-phase interface.  Both designs also feature scintillation photon detection. DUNE expects around 3000~$\nu_e$CC events from a \SI{10}{\kilo\parsec} supernova, providing a $\nu_e$ sensitivity which complements the $\bar{\nu}_e$ ability of most other detectors.  Elastic scattering on electrons and $\bar{\nu}_e$CC events are also expected.  NC nuclear excitations which produce deexcitation gammas are also expected, although the cross sections are currently poorly known. Preliminary studies based on full simulation and reconstruction chains indicate thresholds for efficient reconstruction of between \SIrange{5}{10}{\mega\electronvolt} neutrino energy and energy resolution of around 20\%~\cite{Abi:2020evt, Abi:2020lpk}.   The tracking capability of the TPC enables pointing to the supernova at the several degree level --- see Sec.~\ref{sec:larpoint}.

\subsection{Detection in Other Low-background Detectors}

\subsubsection{The nEXO Experiment}

nEXO is a proposed neutrinoless double-beta decay ($0\nu\beta\beta$) experiment which aims to employ \SI{5}{\tonne} of liquid xenon, enriched to 90\% in the target isotope $^{136}$Xe, inside a cylindrical Time-Projection Chamber (TPC). Both xenon scintillation light (\SI{175}{\nano\meter} wavelength) as well as ionization charge signals will be recorded in the detector using silicon photo-multipliers \cite{Gallina2019} and specialized charge-readout tiles \cite{Jewell2018}, respectively. The simultaneous detection of scintillation and ionization signals allows for reconstruction of an event's energy, position, and multiplicity. The TPC and its cryostat will be located in a \SI{1.5}{\kilo\tonne} water tank instrumented with up to 125 PMTs. These PMTs will detect Cherenkov radiation from traversing muons and allow for a veto to subsequent cosmogenic backgrounds. This muon veto is referred to as the nEXO Outer Detector. Monte Carlo work is being undertaken to optimize PMT placement, develop a trigger scheme, define water-purity requirements, and investigate background contributions from the Outer Detector. Details of the nEXO project and the Outer Detector are described in \cite{nEXO2018}; nEXO's projected sensitivity to $0\nu\beta\beta$ is discussed in \cite{Albert2018}.

While being optimized for the search for $0\nu\beta\beta$, the sizable mass of xenon and water allows for the detection of SN neutrinos from within the galaxy. The \SI{1.5}{\kilo\tonne} of water in the Outer Detector is expected to have a typical response of $\sim$250 neutrino interactions (calculated using the GVKM model \cite{Gava2009}) for a supernova event at a distance of \SI{10}{\kilo\parsec}. The dominant interaction channel will be inverse-beta decay, as expected for a water Cherenkov detector. Supernova neutrinos will also interact with nEXO's xenon volume. The TPC is being optimized to detect $\beta$-like events around the $^{136}$Xe $Q$-value of \SI{2.46}{\mega\electronvolt}. Thus, $\nu_e$ charged current interactions producing excited $^{136}$Cs will be detectable via the nucleus' gamma de-excitation. For supernovae at \SI{2}{\kilo\parsec}, the expected event rate of this interaction channel is  $\sim$5 events over a few seconds. Slightly lower rates are expected for the neutral current inelastic scatters that will produce excited $^{136}$Xe nuclei. However, the majority of supernova-neutrino interactions in the liquid xenon will be through the \cevns{} channel. Here, we expect on the order of 100 supernova neutrino interactions from a GVKM supernova at \SI{10}{\kilo\parsec}, as calculated using methods found in \cite{Lang:2016zhv, XMASS:2016cmy} and cross sections adopted from \cite{Pirinen:2018gsd}. These \cevns{} events are unlikely to trigger data acquisition readout on their own due to the small number of ionization electrons being generated; nevertheless, by employing a buffered readout system and receiving a SNEWS trigger, nEXO could store this data.

\subsection{Detection Estimates}

For reference, we have compiled estimates of the number of detected events in each of the above neutrino detection experiments for three representative models of the neutrino emission from core-collapse supernovae in Tab.~\ref{t:detev}.  The models span the range of the expected diversity in massive stars, however, due to the initial mass function of massive stars, the lower mass models are expected to dominate the population of core-collapse supernovae.  The \SI{11.2}{\solarmass} and \SI{27.0}{\solarmass} presupernova models are taken from \cite{Woosley:2002zz} and modeled in \cite{Mirizzi:2015eza} (the precise models are LS220-s11.2c and LS220-s20.0c and they are taken from the Garching Core-Collapse Supernova Data archive \url{https://wwwmpa.mpa-garching.mpg.de/ccsnarchive/}). The models are simulated until $\sim$\SI{7}{\second} after core bounce. The \SI{40.0}{\solarmass} presupernova models is taken from \cite{Woosley:2007as} and modeled in \cite{OConnor:2014sgn} until black hole formation at $\sim$\SI{540}{\milli\second} after core bounce. Unless otherwise noted, we include estimates for each detector using SNOwGLoBES~\cite{snowglobes} and assume MSW neutrino oscillations via the adiabatic approximation for the normal and inverted mass ordering. All channels included in SNOwGLoBES are included in the total. 

\begin{table}
\begin{tabular}{ccccccc}
Experiment & Type & Mass [kt] & Location & \SI{11.2}{\solarmass} & \SI{27.0}{\solarmass} & \SI{40.0}{\solarmass} \\
{\bf{Super-K}} & H$_2$O/$\bar{\nu}_e$ & 32 & Japan & 4000/4100 & 7800/7600 & 7600/4900 \\
Hyper-K & H$_2$O/$\bar{\nu}_e$ & 220 & Japan & 28K/28K & 53K/52K & 52K/34K \\
{\bf{IceCube}} & String/$\bar{\nu}_e$ & ~2500* & South Pole & 320K/330K & 660K/660K & 820K/630K \\
{\bf{KM3NeT}} & String/$\bar{\nu}_e$ & ~150* & Italy/France & 17K/18K & 37K/38K & 47K/38K \\
{\bf{LVD}} & C$_n$H$_{2n}$/$\bar{\nu}_e$ & 1 & Italy & 190/190 & 360/350 & 340/240 \\
{\bf{KamLAND}} & C$_n$H$_{2n}$/$\bar{\nu}_e$ & 1 & Japan & 190/190 & 360/350 & 340/240 \\
{\bf{Borexino}} & C$_n$H$_{2n}$/$\bar{\nu}_e$ & 0.278 & Italy & 52/52 & 100/97 & 96/65 \\
JUNO & C$_n$H$_{2n}$/$\bar{\nu}_e$ & 20 & China & 3800/3800 & 7200/7000 & 6900/4700 \\
{\bf{SNO+}} & C$_n$H$_{2n}$/$\bar{\nu}_e$ & 0.78 & Canada & 150/150 & 280/270 & 270/180 \\
{\bf{NO{$\nu$}A}} & C$_n$H$_{2n}$/$\bar{\nu}_e$ & 14 & USA & 1900/2000 & 3700/3600 & 3600/2500 \\
{\bf{Baksan}} & C$_n$H$_{2n}$/$\bar{\nu}_e$ & 0.24 & Russia & 45/45 & 86/84 & 82/56 \\
{\bf{HALO}} & Lead/$\nu_e$ & 0.079 & Canada & 4/3 & 9/8 & 9/9 \\
HALO-1kT & Lead/$\nu_e$ & 1 & Italy & 53/47 & 120/100 & 120/120 \\
DUNE & Ar/$\nu_e$ & 40 & USA & 2700/2500 & 5500/5200 & 5800/6000 \\
{\bf{MicroBooNe}} & Ar/$\nu_e$ & 0.09 & USA & 6/5 & 12/11 & 13/13 \\
{\bf{SBND}} & Ar/$\nu_e$ & 0.12 & USA & 8/7 & 16/15 & 17/18 \\
DarkSide-20k & Ar/any $\nu$ & 0.0386 & Italy & - & 250 & - \\
XENONnT & Xe/any $\nu$ & 0.006 & Italy & 56 & 106 & - \\
LZ & Xe/any $\nu$ & 0.007 & USA & 65 & 123 & - \\
PandaX-4T & Xe/any $\nu$ & 0.004 & China & 37 & 70 & - \\
\end{tabular}
\caption{Estimated interaction rates for the detectors (those in operation at the time of writing are bolded) described here for three different models at \SI{10}{\kilo\parsec}, s11.2c and s27.0c from \cite{Mirizzi:2015eza} that form neutron stars and s40 from \cite{OConnor:2014sgn} which forms a black hole.  The two numbers given are the total events over all channels using SNOwGLoBES assuming adiabatic MSW oscillations only for the normal mass ordering (left number) and the inverted mass order (right number). For liquid scintillator experiments, the elastic proton scattering channel is not included; see the individual detector sections for more details on the rates of this interaction. For the string detectors, the mass is given as an effective mass based on 27.0\,$M_\odot$ and the normal mass ordering, for details on the calculation, please see \url{https://doi.org/10.5281/zenodo.4498941}} 
\label{t:detev}
\end{table}
\section{Amateur Astronomer Engagement}
\label{sec:outreach}

The next galactic core-collapse supernova will be a once-in-a-lifetime event. Amateur astronomers will play a vital role in identifying and observing the optical component of the explosion in real time when it appears in the sky. Because SNEWS is designed to provide an early warning of the appearance of such an explosion, the interface between SNEWS and the astronomical community---and particularly amateur astronomers---is critically important. We aim to reinforce these connections and provide the community with information and resources to ensure they are ready to receive SNEWS alerts and participate in the global effort to spot the optical signal. This outreach effort will be guided by three broad thrusts, as described below: awareness, preparedness, and follow-up.

\subsection{Thrusts of Amateur Astronomer Engagement}
\subsubsection*{Thrust 1: Awareness}
Because amateur astronomers will play such an important role in the optical observation of the next galactic supernova, SNEWS intends to strengthen its relationship with the global amateur astronomical community in the coming months and years. The reason is simple: astronomers can only participate if they know that SNEWS exists. This effort will involve reaching out to groups, attending conferences, and engaging with individuals on social media. Cultivating these community connections will be an ongoing process which will support the important goal of maintaining and maximizing observational readiness.

\subsubsection*{Thrust 2: Preparedness}
To prepare the amateur astronomical community to respond quickly and effectively to a neutrino-indicated galactic supernova, it is critical to provide training and guidance regarding how to receive and interpret SNEWS alerts, and what to do when those alerts arrive.

SNEWS maintains a public mailing list that is used to distribute alert notifications in the event of a supernova-like neutrino burst, and anyone can subscribe at the official SNEWS website (\url{https://snews.bnl.gov/}). Alert emails distributed to the list are PGP-signed for authenticity. 
SNEWS also plans to develop additional notification vectors in order to maximize the reach of our alerts to the community. These will likely include utilizing existing multi-messenger networks, such as the Astrophysical Multimessenger Observatory Network (AMON) mentioned in Sec.~\ref{sec:alert-broadcasting}.

In addition to connecting astronomers to the alert system, SNEWS will develop and distribute materials to teach subscribers about the information products available within the alerts and how that information will be updated as subsequent observations are made, so that amateur astronomers are able to easily and quickly identify potential supernova candidates associated with a coincident neutrino burst. 

\subsubsection*{Thrust 3: Follow-up}
When the day finally arrives, the initial SNEWS alert may contain little or no pointing information. It will therefore be up to the professional and amateur astronomy communities to identify and report the precise right ascension and declination of the supernova as it becomes visible. Time will be of the essence; it is imperative that clear and simple reporting mechanisms exist and that observers are aware of them. SNEWS will develop and distribute training materials so that alert subscribers can familiarize themselves with the follow-up reporting process before and during a galactic supernova event.

\subsection{Assessing Observational Readiness}
Galactic core-collapse supernova events are sufficiently rare that most of us will likely only have one opportunity in our lives to observe one. It is therefore of critical importance that the process of disseminating alert information to the astronomical community and facilitating the reporting of real-time follow-up observations be as robust as possible. To characterize this, SNEWS will conduct occasional drills. A drill would involve sending a test alert to SNEWS subscribers which closely resembles a real SNEWS alert and implores subscribers to immediately begin their search for a transient object in the night sky. The transient object would, of course, not be a real supernova, but some other point-like transient source in the sky, and the alert would be clearly labeled as a test. 

In fact, such a drill was once conducted in February of 2003~\cite{Skytelescope2003}. A test alert was distributed to \emph{Sky \& Telescope} AstroAlert subscribers, which provided right ascension and declination coordinates, an uncertainty radius, and a note that the expected magnitude was unknown. Subscribers were encouraged to scan the sky with their eyes, binoculars, and telescopes, and to report any suspected candidates through a web form. The transient test object in this case was the asteroid Vesta, which was successfully identified by six of the 83 submitted reports. 

Each drill will be a valuable case study of our ability to minimize the time between receiving an early sign of an imminent supernova and performing the first optical observations of the explosion. There is much to learn from such a simulation which will allow us to identify potential blind spots and bottlenecks in the process and to mitigate them. For example, the distribution of contributed observations could allow us to identify regions of the world where more engagement between SNEWS and the community is needed. We hope that these drills will also provide valuable feedback to the astronomical community and empower it to identify potential areas for improved training or resource development regarding these types of observations.

SNEWS aims to provide the earliest warning of an imminent galactic supernova, and our ability as a community to identify its precise location in the sky as soon as the optical signal arrives will rely on a global corps of amateur astronomers and their observational expertise. Such timely localization will be vitally important for maximizing the scientific potential of this rare event. Engaging with the amateur astronomer community and providing the training and access to resources necessary to carry out this vital task is essential to the mission of SNEWS.
 \section{Summary}

The SuperNova Early Warning System is one of the oldest multi-messenger astronomy networks, set up to provide advance warning of the next galactic core collapse supernova by forming a coincidence between multiple neutrino detectors around the world.  However, it is also a system which has not yet made a live observation, because such supernovae occur only a few times per century.  Today, multi-messenger astronomy is a burgeoning field across many messengers, and experience gained from successes in simultaneous detection of gamma ray bursts, gravitational waves, and ultra-high energy neutrinos can be applied to creating a new ``SNEWS~2.0'' network which will deliver more neutrino-based information and do so reliably and promptly, to enable the best science possible from the next nearby supernova.
\section{Acknowledgements}

This work is supported by the National Science Foundation ``Windows on the Universe: The Era of Multi-Messenger Astrophysics'' program: ``WoU-MMA: Collaborative Research: A Next-Generation SuperNova Early Warning System for Multimessenger Astronomy'' through grants 1914448, 1914409, 1914447, 1914418, 1914410, 1914416, and 1914426; and via the ``HDR-Harnessing the Data Revolution'' program, grant 1940209. Participation of individual researchers has been supported by funds from the European Union’s Horizon 2020 Research and Innovation Programme under the Marie Sklodowska-Curie grant agreement no.~754496, and research grant number 2017W4HA7S ''NAT-NET: Neutrino and Astroparticle Theory Network'' under the program PRIN 2017 funded by the Italian Ministero dell'Istruzione, dell'Universita' e della Ricerca (MIUR).

\def\newblock{\ } 
\bibliography{bibliography}

\end{document}